\documentclass[11pt,a4paper]{article}
\pdfoutput=1
\usepackage{jheppub}

\def\d{{\textrm d}}

\newcommand{\nn}{\nonumber}

\title{Tachyonic Anti-M2 Branes}

 \author[a]{Iosif Bena,} \author[a]{Mariana Gra\~na,} \author[a]{Stanislav Kuperstein} \author[b]{and Stefano Massai}

 \affiliation[a]{Institut de Physique Th\'eorique, CEA
 Saclay, CNRS URA 2306 \\ F-91191 Gif-sur-Yvette, France} 

 \affiliation[b]{Arnold Sommerfeld Center for Theoretical Physics,\\
  Theresienstr. 37, 80333 Muenchen, Germany} 

\emailAdd{iosif.bena@cea.fr}\emailAdd{mariana.grana@cea.fr}\emailAdd{stanislav.kuperstein@cea.fr}
\emailAdd{stefano.massai@lmu.de}

\abstract{We study the dynamics of anti-M2 branes in a warped Stenzel solution with M2 charges dissolved in fluxes by taking into account their full backreaction on the geometry. The resulting supergravity solution has a singular magnetic four-form flux in the near-brane region. We examine the possible resolution of this singularity via the polarization of anti-M2 branes into M5 branes, and compute the corresponding polarization potential for branes smeared on the finite-size four-sphere at the tip of the Stenzel space. We find that the potential has no minimum. We then use the potential for smeared branes to compute the one corresponding to a stack of localized anti-M2 branes, and use this potential to compute the force between two anti-M2 branes at tip of the Stenzel space. We find that this force, which is zero in the probe approximation, is in fact repulsive! This surprising result points to a tachyonic instability of anti-M2 branes in backgrounds with M2 brane charge dissolved in flux. }

\preprint{IPhT-T14/013, LMU-ASC 04/14}

\begin{document}
\maketitle

\setcounter{footnote}{0}
\setcounter{figure}{0}
\setcounter{equation}{0}

\section{Introduction}

Anti-branes in warped throat geometries are an important ingredient in many models of supersymmetry breaking in string theory. In string phenomenology they represent a generic way of uplifting a given anti-de Sitter (AdS) compactification to a de Sitter (dS) one with small cosmological constant~\cite{Kachru:2003aw}. In holography they are used to construct non-compact flux backgrounds dual to dynamical supersymmetry breaking mechanisms in field theories~\cite{Kachru:2002gs,Argurio:2006ny,Argurio:2007qk}. Recently, anti-branes in flux compactifications have also been used to construct non-extremal black hole microstates~\cite{Bena:2011fc,Bena:2012zi}. Over the past few years there has been an extensive body of work aimed at constructing explicit solutions for the geometry sourced by these anti-branes, and to study their dynamics with full backreaction taken into account.

It is by now well established that if one tries to construct a solution that describes smeared anti-branes placed in a background with positive brane charge dissolved in fluxes by treating the anti-branes as a small perturbation of a supersymmetric solution, one always encounters a singularity coming from a divergent energy density of certain magnetic fluxes. This has been found  for anti-D3 branes in Klebanov-Strassler (KS)~\cite{McGuirk:2009xx,Bena:2009xk,Bena:2011hz,Bena:2011wh}, for anti-M2 branes~\cite{Bena:2010gs,Massai:2011vi,Blaback:2013hqa} in the Cvetic-Gibbons-Lu-Pope (CGLP) solution~\cite{Cvetic:2000db}, as well as for anti-D2 branes in the A8 and CGLP backgrounds~\cite{Giecold:2011gw,Giecold:2013pza,Cottrell:2013asa}. 
Moreover, it has been shown for anti-D3 branes in KS~\cite{Massai:2012jn,Bena:2012bk,Gautason:2013zw} and for anti-D6 branes in a massive type IIA background~\cite{Blaback:2010sj,Blaback:2011nz,Blaback:2011pn,Apruzzi:2013yva} that these singularities are not an artifact of treating the anti-branes as small perturbations, but survive in the fully back-reacted geometry.
Moreover, despite the fact that the singular solution corresponding to smeared anti-D3 branes in KS passes some non-trivial tests~\cite{Bena:2010ze,Dymarsky:2011pm,Dymarsky:2013tna}, it does not appear possible to resolve this singularity by polarizing the anti-D3 branes into D5 branes~\cite{Bena:2012vz}, or by cloaking it with a black hole horizon~\cite{Bena:2012ek,Bena:2013hr,Buchel:2013dla}. Similarly, the aforementioned massive type IIA singularity cannot be cured by polarizing the anti-D6 branes into D8 branes~\cite{Bena:2012tx}.

In this paper we study the solution and the dynamics of fully back-reacted anti-M2 branes in the CGLP background \cite{Cvetic:2000db}. This background is dual to a supersymmetric $\mathcal{N}=2$ (2+1)-dimensional theory obtained by a mass-deformation of the world-volume theory of M2 branes at the tip of a cone over $V_{5,2} = SO(5)/SO(3)$. This field-theory deformation corresponds in supergravity to deforming the cone over $V_{5,2}$ to a Stenzel space, which has a finite-sized $S^4$ at the tip. 

Hence, the CGLP background is the M2-brane analogue of the Klebanov-Strassler solution~\cite{Klebanov:2000hb}. The addition of probe anti-M2 branes to the CGLP geometry has been considered by Klebanov and Pufu~\cite{Klebanov:2010qs} as a way to construct the dual of a long-lived metastable non-supersymmetric state in the field theory. The supergravity solution corresponding to the anti-M2 branes (smeared over the four sphere at the tip of the cone) has been constructed later in~\cite{Bena:2010gs,Massai:2011vi}, by treating the anti-M2 perturbation as a small, first-order deformation of the supersymmetric CGLP background. While this solution has the expected UV properties to correspond to a metastable state, the energy density of the four-form flux diverges in the infrared, near-brane region.

The purpose of this paper is three-fold:

\emph{First}, we want to establish that the singularity of the perturbative solution for anti-M2 branes in the CGLP background does not go away when one constructs the fully back-reacted solution. This is another piece of evidence supporting the idea that the singularities of anti-brane solutions are not artefacts of perturbation theory. 

It is crucial hence to address the question of whether this singularity is physical or not, by searching for an explicit mechanism that can resolve it and this is the \emph{second} purpose of our paper. To be more precise, we examine the possible resolution of this anti-M2 singularity by polarization into M5 branes~\cite{Bena:2000zb}. Klebanov and Pufu have shown in~\cite{Klebanov:2010qs} that probe M2 branes that are localized at the north pole of the $S^4$ at the bottom of the CGLP solution can polarize into M5 branes wrapping an $S^3$ inside this $S^4$. The solutions that we construct have the anti-M2 branes smeared on the $S^4$ in the CGLP infrared, and hence cannot be used to directly determine whether the singularity is cured by this polarization channel.

However, as it is well-known from the extension of the Polchinski-Strassler analysis~\cite{Polchinski:2000uf} to M2 branes~\cite{Bena:2000zb}, these branes can have two polarization channels, corresponding to M5 branes in orthogonal planes. For M2 branes localized on the CGLP infrared $S^4$, these channels correspond to the Klebanov-Pufu M5 brane and to a transverse  M5 brane that  wraps the contractible $S^3$ of the CGLP solution at a finite distance away from the tip. Since this polarization channel is not wiped out by smearing the anti-M2 branes on the $S^4$, we can use our fully-back-reacted solution to calculate its polarization potential. Much like for fully back-reacted anti-D3 branes~\cite{Bena:2012vz}, we will find that the smeared anti-M2 branes do not polarize into this channel. 

The \emph{third} purpose of this paper is to use the polarization potential for the smeared anti-M2 branes in order to calculate that of localized anti-M2 branes, both in the transverse channel as well as into the Klebanov-Pufu (KP) channel.\footnote{As we will see in Section \ref{sec:Localized}, one can relate the smeared and the localized polarization potentials by considering a region in the parameter space where the Schwarzschild radii of the flux and the anti-M2 branes are larger than the radius of the blown-up 4-sphere.} On general grounds, the polarization potential for M2 branes into M5 branes has three terms~\cite{Bena:2000zb}:\footnote{Throughout the Introduction, the coordinate $r$ will denote the coordinate distance from the brane sources.} one proportional to $r^6$, which is always positive, one proportional to $r^4$, which is negative and which comes from the flux that forces the polarization to happen, and one proportional to $r^2$, which is the same as the potential felt by mobile M2 branes in the background. This latter contribution is zero if one considers the anti-M2 branes as probes moving on the $S^4$ in the infrared of CGLP \cite{Klebanov:2010qs}, and this reflects the fact that there is no preferred position for these anti-M2 branes on the $S^4$ due to the space isometry. However, once one places a stack of (back-reacted) anti-M2 branes at a given point on the $S^4$ this symmetry is broken and one expects other anti-M2 branes to feel a non-trivial force. 

To compute the polarization potential for the KP channel one first needs  to realize that the transverse polarization potential is the sum of two contributions, which have very different holographic origins: one term comes from giving a supersymmetric mass to the fermions on the (anti) M2 brane world-volume and to their bosonic partners, and is a perfect square. The second contribution comes from traceless boson bilinears (and can therefore be called an  $L=2$ contribution)~\cite{Bena:2000zb}. This contribution can in principle be given by any traceless symmetric $8 \times 8$ matrix, $m_{ij}$, sandwiched between the eight scalars, $\phi^i$, of the M2 branes ($\phi^i m_{ij} \phi^j$), but when the anti-M2 branes are all localized at one point on the $S^4$, symmetry dictates that only two such terms can exist, and only one of the two is relevant for the polarization potential. This allows us in turn to disentangle the ``susy''  and the $L=2$ contributions to the transverse polarization potential, and to use them to reconstruct the polarization potential for the KP channel. A striking surprise awaits: this potential has an $r^2$ term that is the negative of a perfect square. Thus, if one places two stacks of anti-M2 branes at the bottom of the CGLP background, the force between these two stacks is always repulsive, independently of the parameters that determine the solution asymptotically. Hence, the theory on the world-volume of these anti-M2 branes is \emph{tachyonic}!

This result, which contradicts the expectations one might have formed by naively extrapolating the ``giant inflaton'' arguments of \cite{DeWolfe:2004qx},\footnote{For more details see Section \ref{sec:concl}.}  has several unexpected consequences. 

First, it implies that the singularity of the localized anti-M2 brane solution~\cite{Blaback:2013hqa} is worse than one might have thought. Indeed, if one imagines putting together many anti-M2 branes and holding them by force, one can expect that these anti-M2 branes will develop an $AdS_4 \times S^7$ throat, perhaps perturbed with some fluxes. Our result shows that other anti-M2 branes placed in this throat are repelled towards its UV, and hence this throat is unstable to fragmentation.

Second, if one imagines placing a stack of anti-M2 branes inside the CGLP solution, the world-volume theory on these anti-M2 branes develops a tachyon. Note that this tachyon cannot be seen in the first-order perturbative description of the anti-M2 branes that uses their Born-Infeld-like brane action. This tachyon rather comes from terms in this action that are quadratic in the transverse magnetic fields, which the Born-Infeld action does not see. 

Third, this negative mass also has the potential to destabilize the metastable minimum found in \cite{Klebanov:2010qs} in the probe analysis. Indeed, the infrared expansion of the M5 potential in the probe limit only has $r^6$ and $r^4$ terms, and the existence of a metastable vacuum comes from the interplay between these terms and the curvature of the $S^4$. Adding another negative term in the game can completely wipe out this metastable vacuum. However, as we will discuss in Section \ref{sec:concl}, the tachyon will not destroy the vacua that correspond to polarizing the anti-M2 branes into multiple M5 branes, although it will probably introduce new instabilities for these vacua.

The paper is organized as follows. In Section~\ref{sec:Setup} we present the supersymmetric solutions corresponding to self-dual (anti-self dual) fluxes and M2 (${\overline{\rm M2}}$) branes. In Section~\ref{sec:interpolate} we show that the solution that interpolates between the CGLP ultraviolet and smeared anti-M2 branes in the infrared is singular. In Section~\ref{sec:generalapproach} we discuss the basic features of brane polarization in asymptotically $AdS$ geometries. In Section~\ref{subsec:antiM2cglp} we compute the polarization potential for smeared anti-M2  branes in the CGLP background. In Section~\ref{sec:Localized} we extend the calculation to find the potential for localized sources. In Section~\ref{sec:validity} we discuss in detail the approximations used, and we conclude in Section~\ref{sec:concl}. Further technical details and discussions are left to the appendices.

\section{Supergravity solutions on a Stenzel space}
\label{sec:Setup}

In this section we start with a short review of the supersymmetric flux solution first constructed by Cvetic-Gibbons-Lu-Pope (CGLP) in~\cite{Cvetic:2000db}, based on a warped Stenzel space~\cite{Stenzel:1993}. We also discuss the most general supersymmetric solutions with self-dual (SD) and anti-self-dual (ASD) fluxes on Stenzel background and show that only the former admits a regular solution.

\subsection{Stenzel Ansatz}
\label{subsec:sugraansatz}

\begin{figure}[t]
\centering
\includegraphics[scale=0.3]{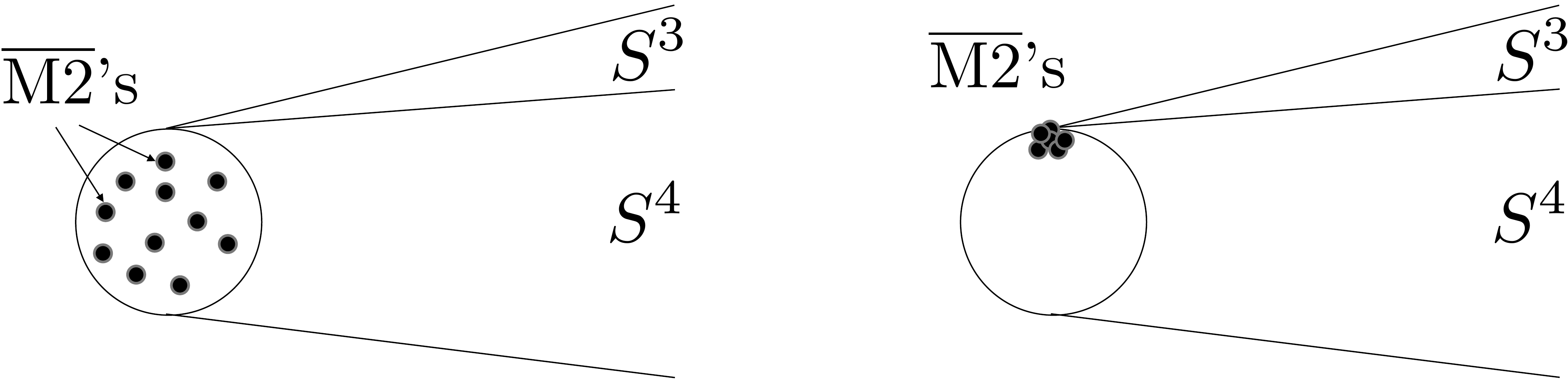}
\caption{Anti-M2 branes smeared over (left) and localized at a point (right) on the 4-sphere at the apex of the 8-dimensional Stenzel space.}
\label{fig:Smeared-Localized}
\end{figure}

Let us start with a presentation of the $11d$ supergravity Ansatz of~\cite{Cvetic:2000db}. It describes the most general M2-like solution that preserves both the $SO(1,2)$ Poincar\'e symmetry of the M2 brane world-volume and the isometry of the internal Stenzel space. This solution describes both M2 brane charge dissolved in the fluxes, as well as M2-brane sources smeared on the $S^4$ at the tip of the Stenzel space. The difference between  smeared and localized M2 sources is illustrated in Figure \ref{fig:Smeared-Localized}. In this section and the following ones we will only analyze smeared sources; those interested in the physics of localized sources will have to wait until Section~\ref{sec:Localized}.

One can parametrize the eleven-dimensional metric in the familiar M2-form:
\begin{equation}
\label{11d-metric}
\d s_{11}^2 = e^{-2 z} \d x_\mu \d x^\mu + e^z \d s_{8}^2 \, ,
\end{equation}
where the 8-dimensional metric has the most general structure consistent with the isometries of the original Stenzel metric:
\begin{equation}
\label{Stenzel-metric}
\d s_{8}^2 = e^{2 \gamma} \left( \d \rho^2 + \nu^2 \right) + e^{2 \beta} \sum_{i=1}^3 \widetilde{\sigma}_i^2 + e^{2 \alpha} \sum_{i=1}^3 \sigma_i^2   \, .
\end{equation}   
We refer the reader to~\cite{Cvetic:2000db} for the definitions of the seven angular one-forms. The functions $\alpha$, $\beta$, $\gamma$, as well as the 11$d$ warp function $e^z$ in \eqref{11d-metric}, depend only on the radial coordinate $\rho$. For the deformed Stenzel space the 3-cycle spanned by $\widetilde{\sigma}_i$ shrinks at the apex, while the 4-cycle corresponding to the remaining four 1-forms attains a fixed size (we will return to this issue later in the paper). 
The Ansatz for the 4-form flux is:
\begin{eqnarray}
\label{4-form}
G_4 &=& \d K \wedge \d x_0 \wedge \d x_1 \wedge \d x_2  + F_4 \, , \qquad \textrm{where} \,  
\nonumber \\
F_4 &=&  \d \left( f \cdot \widetilde{\sigma}_1 \wedge \widetilde{\sigma}_2 \wedge
\widetilde{\sigma}_3 + h \cdot \epsilon^{ijk} \sigma_i \wedge \sigma_j
\wedge \widetilde{\sigma}_k \right) =
\\
&=& f^\prime \cdot \d \rho \wedge \widetilde{\sigma}_1 \wedge \widetilde{\sigma}_2 \wedge
\widetilde{\sigma}_3 + h^\prime \cdot \epsilon^{ijk} \d \rho \wedge \sigma_i \wedge \sigma_j
\wedge \widetilde{\sigma}_k
\nonumber  \\
&& \quad + \frac{1}{2} (4h-f) \cdot \epsilon^{ijk} \nu \wedge \sigma_i \wedge
\widetilde{\sigma}_j \wedge \widetilde{\sigma}_k  -6 h \cdot \nu \wedge \sigma_1 \wedge \sigma_2 \wedge \sigma_3 \, . 
\nonumber 
\end{eqnarray}
Here $f$, $h$ and $K$ are all functions of $\rho$ and ${}^\prime$ denotes the $\rho$-derivative. As usual, we will refer to the component of the 4-form flux $G_4$ in \eqref{4-form} that is extended along the time direction as the electric component, and to $F_4$ as the magnetic component. Departing from the conventions adopted in \cite{Cvetic:2000db} and in follow-up papers, we will omit an overall factor of $m$ in the definition of $F_4$ by absorbing it in $f$ and $h$.

An explicit relation between $K(\rho)$ and the functions appearing in the form $F_4$, $f(\rho)$ and $h(\rho)$, can be derived using the $G_4$ equation of motion:
\begin{equation}
\label{G4-EOM}
\d \star_{11} G_4 = - \frac{1}{2} G_4 \wedge G_4 \, .
\end{equation}
We get:\footnote{The $8d$ space orientation is given by
\begin{equation}
\star_8 \d \rho = e^{3(\alpha+\beta)} \nu \wedge \sigma_1 \wedge \sigma_2 \wedge \sigma_3 \wedge \widetilde{\sigma}_1 \wedge \widetilde{\sigma}_2 \wedge
\widetilde{\sigma}_3 \, ,
\end{equation}
and for the $11d$ orientation we set:
\begin{equation}
\star_{11} F_4 = e^{-3 z} \star_8 F_4 \wedge \d x_0 \wedge \d x_1 \wedge \d x_2 \, .
\end{equation}
With these conventions the flux will be self-dual for a supersymmetric solution with mobile M2's and anti self-dual for anti-M2's (see later on).  
}
\begin{equation}
\label{K-prime}
K^\prime = 6 e^{-3 \left( \alpha+\beta+2 z \right)} \left( h (f - 2 h) - P \right) \, .
\end{equation}
Here $P$ is an integration constant related (but necessarily not proportional) to possible brane sources as we will review shortly.

In this paper we are interested in the most general solution of the form \eqref{11d-metric},  \eqref{Stenzel-metric} and \eqref{4-form}. Integrating over the angles and the space-time coordinates, the $11d$ supergravity action becomes (up to an overall factor) a functional, $\mathcal{L}= - \frac{1}{2} G^{ab}(\phi) {\phi_a}^\prime {\phi_b}^\prime-V(\phi) $, of six $\rho$-dependent functions: 
\begin{equation}
\phi^a(\rho) = \big( \alpha(\rho),\beta(\rho),\gamma(\rho),z(\rho),f(\rho),h(\rho) \big) \, .
\end{equation}
The kinetic term of this Lagrangian is \cite{Cvetic:2000db}:
\begin{eqnarray}
\label{G-metric}
G_{a b}\, {\phi^a}^\prime\, {\phi^b}^\prime &=& - 6  e^{3(\alpha+\beta)} \left( {\alpha^\prime}^2 + 3 {\alpha^\prime} {\beta^\prime} + {\beta^\prime}^2 + \left( {\alpha^\prime} + {\beta^\prime} \right) {\gamma^\prime} -\dfrac{3}{4} {z^\prime}^2 \right) \\
\nonumber
&& \qquad \qquad \qquad \qquad
 + \dfrac{1}{2} e^{- \alpha - 3(\beta+z)} \left( e^{4\alpha} {f^\prime}^2 + 12 e^{4\beta} {h^\prime}^2 \right) \, ,
\end{eqnarray}
while the potential terms come from a superpotential $W$ via:
\begin{equation}
 V(\phi) = \frac{1}{8}\, G^{ab}\, \dfrac{\partial W}{ \partial \phi^a}\, \dfrac{\partial W}{ \partial \phi^b} \, ,
\end{equation}
with two possible solutions for $W$:
\begin{equation}
\label{superpotential}
W(\phi)= - 3 e^{2 \left( \alpha + \beta \right)}  \left(  e^{2 \alpha} + e^{2 \beta} + e^{2 \gamma} \right) \mp 6 e^{-3 z} \left( h(f-2h) - P \right) \, .
\end{equation} 
The fact that two different superpotentials (with different signs in \eqref{superpotential}) reproduce the same potential tells us that there are two possible supersymmetric solutions of the EOMs: one with mobile M2's and self-dual (SD) 4-form flux, and the other with mobile anti-M2's  and anti-self-dual (ASD) flux.  Notice that the constant $P$ should therefore be different for the two possibilities. We will denote the ``$-$" option in \eqref{superpotential} by $W_\textrm{SD}$, and the ``+" by $W_\textrm{ASD}$. The two solutions are (potentially) supersymmetric, but preserve different sets of supercharges.\footnote{Strictly speaking, there is a possibility that the first-order equations correspond only to fake supersymmetry, like in \cite{Kuperstein:2003yt,Halmagyi:2011yd}, but this is irrelevant to our discussion.}
 

Let us provide more details about the solutions derived from the two superpotentials $W_\textrm{SD}$ and $W_\textrm{ASD}$. By using a standard and useful notation (see for example~\cite{Borokhov:2002fm,Bena:2012vz}) we introduce a set of six functions $\xi_a$ ($a=\alpha, \beta, \dots, h$) dual to the modes $\phi^a$, such that the first-order equations coming from supersymmetry can be written in the following general form:
\begin{equation}
\label{FirstOrderSUSY}
\xi_a = 0 
\qquad \textrm{where} \qquad
    \xi_a \equiv G_{a b} {\phi^b}^\prime - \frac{1}{2} \frac{ \partial W}{ \partial \phi^a} \, .
\end{equation}
In what follows we will use the obvious notations $\xi_a^+$ and $\xi_a^-$ for $\xi_a$'s defined with $W=W_\textrm{SD}$ and $W=W_\textrm{ASD}$ respectively. We present the explicit form of these functions for the superpotentials (\ref{superpotential}) and the metric (\ref{G-metric}) in Appendix~\ref{sec:xi's}.

The functions $\xi_a$'s defined in (\ref{FirstOrderSUSY}) do not vanish for a general non-supersymmetric solution. In fact they satisfy an additional set of first-order ODEs ~\cite{Bena:2012bk,Bena:2012vz}:
\begin{equation}
\label{SecondOrder}
    {\xi_a}^\prime = - \frac{1}{2} \left[ \frac{\partial G^{b c}}{\partial \phi^a} \xi_b \xi_c
                                   + \frac{\partial G^{b c}}{\partial \phi^a} \frac{\partial W}{\partial \phi^b} \xi_c
                                   + G^{b c} \frac{\partial^2 W}{\partial {\phi^a} \partial {\phi^b}} \xi_c
                                   \right] \, ,
\end{equation}
which is trivially solved by $\xi_a=0$ for a supersymmetric solution. Throughout this paper we will only need three equations, one for the warp function and two for the flux:
\begin{eqnarray}
\label{xi-fhz-dot}
{\xi^-_f}^\prime &=& 2 e^{-3 (\alpha+\beta+z)} h \, \xi^-_z - \frac{1}{2} e^{\alpha-\beta}  \xi_h^- \, ,
\nonumber \\
{\xi^-_h}^\prime &=& 2 e^{-3 (\alpha+\beta+z)} (f-4h) \, \xi^-_z - 6 e^{3(\beta-\alpha)}  \xi_f^- + 2 e^{\alpha-\beta}  \xi_h^- \, ,
\\
{\xi^-_z}^\prime &=& - K^\prime e^{3 z} \xi^-_z - \frac{e^{3 z}}{4} \left( 12 e^{3 (\beta-\alpha)} {\xi_f^-}^2  +  e^{\alpha-\beta} {\xi_h^-}^2 \right)  
 \nonumber \, .
\end{eqnarray}
The explicit form of the remaining equations for our Ansatz is relegated to Appendix~\ref{sec:xi's-dot}. We will return to the (non-supersymmetric) second-order equations of motion later in this section.

\subsection{Solutions with SD and ASD fluxes}
\label{subsec:susyCGLP}

In this subsection we would like to study the most general solutions emerging from the set of eight first-order equations in (\ref{FirstOrderSUSY}). We will also comment on the charges of these solutions.

Clearly, for $\phi^a=\alpha, \beta$ or $\gamma$ we have $\xi_a^+=\xi_a^-$, since these fields do not appear in the flux part of the superpotential \eqref{superpotential}. Solving the three first-order ODEs, $\xi_a=0$, for these metric functions leads to the Stenzel metric (see (2.31) of \cite{Cvetic:2000db}). In doing so, one has to fix three integration constants. One constant is related to the possible shift $\rho \to \rho + \textrm{const}$ (the only remnant of the $\rho$-reparametrization invariance of the Lagrangian), other constant has to take a specific value\footnote{We disagree on this point with \cite{Cvetic:2000db} where this constant is fixed ``without loss of generality".} in order to avoid a singularity for small $\rho$ and, finally, the third constant corresponds to an arbitrary overall rescaling of the $8d$ metric. In our notations the final result is:
\begin{eqnarray}
\label{Stenzel-functions}
e^{2 \bar{\alpha}} &=& \dfrac{3^{3/4}}{2} \epsilon^{3/2} \left( 2 +  \cosh(2 \rho) \right)^{1/4} \cosh \rho
\nonumber  \\
e^{2 \bar{\beta}} &=& \dfrac{3^{3/4}}{2} \epsilon^{3/2} \left( 2 +  \cosh(2 \rho) \right)^{1/4} \dfrac{\sinh^2 \rho}{\cosh \rho} \\
e^{2 \bar{\gamma}} &=& \dfrac{3^{7/4}}{2} \epsilon^{3/2} \left( 2 +  \cosh(2 \rho) \right)^{-3/4} \cosh^3 \rho 
\, ,
\nonumber
\end{eqnarray}
where the bar in $\bar \alpha, \bar \beta$ and $\bar \gamma$ stands for the background (or GCLP) value and $\epsilon$ is the deformation parameter measuring the size of the blown up $S^4$ at $\rho=0$.\footnote{In the notations of \cite{Cvetic:2000db} one has $\epsilon=2^\frac{5}{6}/3^{\frac{2}{3}}$ and for \cite{Klebanov:2010qs} the identification is $\epsilon=2/3^{\frac{7}{4}}$.} For large $\rho$ all of the functions in \eqref{Stenzel-functions} behave as $e^{\frac{3}{2} \rho} \sim r^2$, where $r$ is the radial coordinate with which the metric on the singular, $\epsilon=0$, Stenzel space (or alternatively far away from the deformed apex) takes an explicit conic form $\d r^2 + r^2 \d s_{V(5,2)}^2$.

\subsubsection{Self-dual flux and M2-branes}

Contrary to the metric $\xi$'s, the three remaining $\xi$  functions depend on the choice of sign in (\ref{superpotential}). As we have already mentioned earlier, for $W_\textrm{SD}$ and $W_\textrm{ASD}$ the 4-form flux $F_4$ in (\ref{4-form}) has to be self- and anti-self-dual respectively. In other words, the equations $\xi_f^+=0$ and $\xi_h^+=0$ are equivalent to $F_4 = \star_8 F_4$ in our notations. The most general solution is:
\begin{eqnarray}
\label{SDflux}
f(\rho) &=& C_1 \dfrac{3 \cosh^2 \rho - 1}{\cosh^3 \rho} + C_2 \cosh \rho \left( \cosh^2 \rho - 3 \right) \\
\nonumber
h(\rho) &=& \dfrac{C_1}{2 \cosh \rho} - \dfrac{C_2}{2} \cosh^3 \rho \, .
\end{eqnarray}
For the UV asymptotics to be that of the regular M2 background (meaning $e^{-\frac{9}{2} \rho} \sim r^{-6}$ decay of the warp function), one has to set $C_2=0$. Adhering to the conventions of \cite{Cvetic:2000db} we define the remaining constant as:
\begin{equation}
\label{C1}
C_1 = - \dfrac{\sqrt{3}}{9} M \, ,
\end{equation}
where the flux parameter $M$ is the same as  $m$ in \cite{Cvetic:2000db} and the follow-up papers.

The magnetic flux $F_4$ and $M$ are related by (see for example (42) of \cite{Klebanov:2010qs}):
\begin{equation}
\label{M5-charge}
\dfrac{8 \pi^2}{3 \sqrt {3}} \cdot \dfrac{M}{\left( 2 \pi l_\textrm{P}\right)^3} = \dfrac{1}{\left( 2 \pi l_\textrm{P}\right)^3} \int_{S^4} F_4  \equiv \widetilde{M} \, ,
\end{equation}
where $\widetilde{M}$ is a dimensionless quantity used in \cite{Klebanov:2010qs}, which Dirac quantization fixes to be an integer. The only reason we use the non-integer $M$ in this paper is because it does not appear explicitly in (\ref{4-form}), but rather directly in the flux functions $h$ and $f$.\footnote{Note also that the parameter $m$ used in \cite{Klebanov:2010qs} is different from the one used here and in \cite{Cvetic:2000db}:
$$
\frac{27 \sqrt{3}}{4} m_\textrm{\cite{Klebanov:2010qs}} =  m_\textrm{\cite{Cvetic:2000db}} = M \, .
$$
}

The Maxwell charge of the 4-form (\ref{4-form}) is:\footnote{In deriving this result one might use:
$$
\int \nu \wedge \sigma_1 \wedge \sigma_2 \wedge \sigma_3  \wedge \widetilde{\sigma}_1 \wedge \widetilde{\sigma}_2 \wedge \widetilde{\sigma}_3 = \dfrac{16}{3} \pi^4 \, .
$$
This result follows from the asymptotic form of Stenzel metric (see, for instance, (14) of \cite{Pufu:2010ie}) and the fact that $\textrm{Vol}\left( V_{5,2} \right) = 27 \pi^4/128$, as was originally derived in \cite{Bergman:2001qi}. 
Notice also that the $32 \pi^4$ numerical factor in (\ref{Maxwell}) is different from the one in \cite{Bena:2010gs}, but matches all other references.}
\begin{equation}
\label{Maxwell}
Q^\textrm{Maxwell}_\textrm{M2} (\rho ) = \dfrac{1}{\left( 2 \pi l_\textrm{P}\right)^6}  \int\displaylimits_{V_{5,2}, \,\, \rho = \textrm{const}} \star_{11} G_4 = \dfrac{32 \pi^4}{\left( 2 \pi l_\textrm{P}\right)^6}  \Big( P - h(\rho) \left( f(\rho) - 2 h(\rho) \right) \Big) \, .
\end{equation} 
Since at $\rho=0$ the space is perfectly smooth, the only possible contribution to the Maxwell charge comes from the M2 sources smeared over the $S^4$ at the (blown up) tip. Denoting the number of the M2 sources by $N_\textrm{M2}$ and reading $f(0)$ and $h(0)$ from (\ref{SDflux}) and (\ref{C1}) we get:
\begin{equation}
\label{P}
P = \frac{M^2}{54} + \dfrac{\left( 2 \pi l_\textrm{P}\right)^6}{32 \pi^4} N_\textrm{M2} \, .
\end{equation} 
With this assignment for $P$, the asymptotic values of the Maxwell charge are:
\begin{equation}
\label{Maxwell-asympt}
Q^\textrm{Maxwell}_\textrm{M2} (0) = N_\textrm{M2}
\qquad \textrm{and} \qquad
Q^\textrm{Maxwell}_\textrm{M2} (\infty) = \dfrac{\widetilde{M}^2}{4} + N_\textrm{M2} \, .
\end{equation}
For $N_\textrm{M2}=0$ these results were first observed in \cite{Martelli:2009ga}. It was also noted there that since the UV Maxwell charge has to be integer, $\widetilde{M}$ (defined in (\ref{M5-charge})) is necessarily even.\footnote{In general, the Maxwell charge, though conserved, is not quantized. It interpolates smoothly between the two integer asymptotic values. The fully detailed analysis of the Maxwell, brane and (quantized) Page charges in Stenzel geometry appears in \cite{Hashimoto:2011aj} and \cite{Hashimoto:2011nn}.}

\subsubsection{Anti self-dual flux and anti-M2 branes} 
\label{subsubsec:ASD}

Anti-self dual flux, $F_4 = - \star_8 F_4$, is obtained from requiring $\xi_f^-=0$ and $\xi_h^-=0$. This time the general solution is:
\begin{eqnarray}
\label{anti-SDflux}
f(\rho) &=& \dfrac{2}{\cosh^3 \rho} \left( -\widetilde{C}_1 + \widetilde{C}_2 \left( 3 \cosh^4 \rho + 1 \right) \right) \\
\nonumber
h(\rho) &=& \dfrac{1}{\cosh \rho} \left( \dfrac{\widetilde{C}_1}{\sinh^2 \rho} + \widetilde{C}_2  \left( \cosh^2 \rho + 1 \right) \right) \, .
\end{eqnarray}
Regularity in the IR requires $\widetilde{C}_1=0$, as otherwise both the flux blows up and the resulting warp function $e^{3 z}$ behaves as $\rho^{-6}$ for small $\rho$ leading to a naked singularity at $\rho=0$. The $\widetilde{C}_2$ mode destroys instead the UV asymptotics. For large $\rho$, the $\xi^-_z=0$ equation (see (\ref{xi-zfh})) implies that $e^{3 z} \sim e^{-\frac{5}{2} \rho}$, in contrast to the asymptotically-$AdS$ solution which has $e^{3 z} \sim e^{-\frac{9}{2} \rho}$. With both types of fluxes, the UV- or the IR-divergent one,  we can add mobile anti M2's at the tip, though this obviously will not cure the corresponding singularities.  

What we find is conceptually different from the type IIB conifold-based story \cite{Klebanov:2000hb}. There, the solution with imaginary anti-self dual flux\footnote{\label{G_3} Satisfying $\star_6 G_3 = -i  G_3$, where the complex 3-form flux is defined as $G_3 \equiv F_3 - i g_S^{-1} H_3$.} is not really different from its self-dual counter-part. Instead, the two solutions are trivially related by the sign flip of the $B$-field, $B_2 \to -B_2$. The two solutions are everywhere regular and supersymmetric, though, they preserve different supercharges. 

The Stenzel space does not have an analogous $\mathbb{Z}_2$ symmetry and as a consequence the SD and the ASD flux equations produce completely different results. In particular, both ASD solutions turn out to be singular, either in the IR or the UV. This aspect of Stenzel ASD fluxes has been overlooked in the literature.

\section{Absence of a regular solution with anti-M2 branes and asymptotic self-dual flux}
\label{sec:interpolate}

In this section we would like to demonstrate that there is no solution of the second-order equations of motion with regular fluxes that interpolates between the following two asymptotic solutions derived from the superpotential (\ref{superpotential}):
\begin{itemize}
\item Anti-M2's smeared over the 4-sphere at the tip with some amount of IR regular anti-self dual (ASD) flux, described by Eq. (\ref{anti-SDflux}) with $\widetilde{C}_1=0$ (plus some amount of IR regular self-dual flux) and 
\item The CGLP background with $M$ units of self-dual (SD) flux (described by Eq. (\ref{SDflux}) with $C_2=0$) in the UV.
\end{itemize}
A similar calculation has been carried out in \cite{Bena:2012bk} for anti-D3's in the Klebanov-Strassler geometry, where it was shown that starting with anti-D3 smeared at the KS tip one ends up with IASD flux all the way to the UV, unless the flux is allowed to be singular in the IR.

The output of this section is strictly speaking of no immediate importance for our main conclusions in the paper. The reader can skip this section without losing the thread of the upcoming arguments.

We will pursue the following strategy. We will first write down the flux regularity conditions, and then will use them to identify the lowest $\rho$ powers in the Taylor expansions of the $\xi^-$ functions. Plugging this into the ${\xi^-_{z,f,h}}^\prime$ equations we will finally argue that $\xi^-_f$ and $\xi^-_h$ vanish identically, implying that the flux remains ASD all the way to the UV.

To proceed we have to elaborate first on the near-apex (small $\rho$) behavior of the $8d$ metric. For a general $\mathcal{N}=0$ solution one cannot rule out any leading order terms in the expansion of the functions $\alpha$, $\beta$ and $\gamma$ in (\ref{Stenzel-metric}). We, however, do not want to ruin the topological structure of Stenzel space (\ref{Stenzel-metric}). In order words, independently of the fluxes and the source backreaction, the $8d$ metric should describe a regular space at the tip, since otherwise the source (and therefore the brane) interpretation  will be, strictly speaking, meaningless. In practical terms it means that the 3-sphere should shrink smoothly at $\rho=0$. A simple calculation shows that it happens if and only if both $e^\alpha$ and $e^\gamma$ approach constant values at $\rho=0$, while $e^{\beta - \gamma} = \rho + \ldots$ at small $\rho$. Throughout the paper we will insist on this behavior. There are no further restrictions on the three functions. In particular, the sizes of the 4-sphere and of the $U(1)$ fibre in $\nu$ are free parameters.\footnote{We believe that the regularity of the $8d$ metric at $\rho=0$ can be consistently derived from the supergravity equations of motion, although the full analysis appears to be difficult.}  

Next, the warp factor $e^{3 z}$ should behave like $\rho^{-2}$ near the tip, where the coefficient of proportionality is fixed by the number of smeared anti-M2's. If this does not happen we cannot interpret the small-$\rho$ region of the geometry as having anything to do with the backreaction of smeared anti-M2 branes. Note that in this section we insist on this behavior for the warp factor and on having no divergent magnetic flux. In Section~\ref{subsec:antiM2cglp} we will relax these assumptions. 

We now can derive the conditions for flux regularity. A straightforward calculation leads to the following results for the IR behavior of the magnetic flux density:\footnote{For simplicity we omit here the subscript indices of the 1-forms $\sigma_{i=1,2,3}$ and $\widetilde{\sigma}_{i=1,2,3}$.} 
\begin{eqnarray}
&&
F_{\rho \widetilde{\sigma} \widetilde{\sigma} \widetilde{\sigma}} F^{\rho \widetilde{\sigma} \widetilde{\sigma} \widetilde{\sigma}} \sim {f^\prime}^2 e^{-4 z - 6 \beta - 2\gamma}
\qquad
F_{\rho \sigma \sigma \widetilde{\sigma}} F^{\rho \sigma \sigma \widetilde{\sigma}} \sim {h^\prime}^2 e^{-4 z - 4 \alpha  - 2 \beta - 2 \gamma}
\\
&&
\quad
F_{\nu \sigma \widetilde{\sigma} \widetilde{\sigma}} F^{\nu \sigma \widetilde{\sigma} \widetilde{\sigma}} \sim \left( f - 4 h \right)^2 e^{-4 z - 2 \alpha - 4 \beta - 2\gamma}
\qquad
F_{\nu \sigma \sigma \sigma} F^{\nu \sigma \sigma \sigma} \sim h^2 e^{-4 z - 6 \alpha - 2 \gamma} \, .
\nonumber
\end{eqnarray}
Thus the IR regularity implies that the Taylor expansions of the functions $f(\rho)$ and  $h(\rho)$ have to be of the form:
\begin{equation}
\label{fh-RegIR}
f (\rho) = 4 h_{(0)} + f_{(3)} \cdot \rho^3 + \ldots
\qquad \textrm{and} \qquad
h (\rho) = h_{(0)} + h_{(1)} \cdot \rho + \ldots \, .
\end{equation}
From this it follows that $K^\prime e^{3 z} = \mathcal{O} \left( \rho^{-1} \right) $ and we can also derive the restrictions on the IR Taylor expansions of the functions $\xi_z^-$, $\xi_f^-$ and $\xi_h^-$. Let us denote by $n_z$, $n_f$ and $n_h$ the lowest powers in the expansions. Then:
\begin{equation}
\label{xi-zfhRegIR}
n_z \geqslant 2 \, , \qquad
n_f \geqslant 1 \, , \qquad
n_h \geqslant 3 \, .
\end{equation}

The last piece of information that we need is the behavior of the metric functions for small $\rho$. As we just pointed out, there is no topological restriction on the (constant) value of $e^{\alpha(\rho)}$ at $\rho=0$. The equations of motion, however, imply that $e^{\beta-\alpha}=\rho+\mathcal{O} \left( \rho^2 \right)$. Let us show how this works: Using the restrictions on the metric and warp functions we just described, one can demonstrate that all three functions $\xi_\alpha^-$, $\xi_\beta^-$ and $\xi_\gamma^-$ start with $\rho^2$.
We will introduce the notation $e^{\beta-\alpha}=t \cdot \rho+\ldots$, where $t$ is the constant we are interested in. The ${\xi_\gamma^-}^\prime$ equation (the only one in (\ref{xi-abg}) with no flux functions involved) implies that $t=1$ or $t=5/7$. On the other hand, the ${\xi_\alpha^-}^\prime- {\xi_\beta^-}^\prime$ holds if and only if $t=1$. To arrive at this result it is important to notice that by virtue of (\ref{fh-RegIR}) and (\ref{xi-zfhRegIR}) the flux functions in this equation can contribute only at the $\rho^3$ order. One may further show that $t=1$ is also consistent with the remaining  ${\xi_\alpha^-}^\prime + {\xi_\beta^-}^\prime$ equation.

We are now in a position to demonstrate that starting from (\ref{xi-zfhRegIR}) one finds only the trivial $\xi^-_{f}=\xi^-_{h}=0$ solution. In other words, the flux of the solution with nonsingular infrared will remain ASD all the way to the UV and can never become SD as in the CGLP background. To show this we need to use equations (\ref{xi-fhz-dot}). The last equation holds only if one of the two conditions is satisfied:
\begin{equation} \label{orders}
n_z - 1 = 2 n_f +1 \geqslant 2 n_h - 3 
\qquad \textrm{or} \qquad
n_z - 1 = 2 n_h - 3\geqslant  2 n_f +1
\end{equation}
Carefully inspecting the equations for ${\xi_f^-}^\prime$ and ${\xi_h^-}^\prime$ and using (\ref{orders}), we observe that $\xi_z^-$ is subleading in both equations. For small $\rho$ we get:
\begin{equation}
 {\xi_f^-}^\prime = - \frac{1}{2 \rho}  \xi_h^- + ...
\, , \qquad \qquad
 {\xi_h^-}^\prime = - 6 \rho^3  \xi_f^- +  \dfrac{2}{\rho} \xi_h^- +... \, .
\end{equation}
where $+ \dots$ stands for higher order terms. Solving this we arrive at $\xi_f^- \sim \rho^2$, $\xi_h^-\sim \rho^2$. The latter is however in  contradiction with (\ref{xi-zfhRegIR}). We conclude that both $\xi_f^-$ and $\xi_h^-$ have to be zero.

We conclude, therefore, that there is no solution with a non-singular infrared flux that interpolates between a solution with smeared anti-M2's and ASD flux in the IR and the SD background of CGLP in the UV.

\section{Basics of brane polarization}
\label{sec:generalapproach}

The main problem that we will address in the next sections is the study of the dynamics of anti-M2 branes at the tip of the CGLP geometry, with the purpose of checking whether the singular anti-M2 solution is resolved by brane polarization, as suggested by the probe analysis of~\cite{Klebanov:2010qs}. 

Before doing this, we review in this section some basic aspects of brane polarization in curved spacetimes, with the purpose of collecting a number of known facts that will be crucial when we will address the polarization of anti-M2 branes in the CGLP background in Section~\ref{subsec:antiM2cglp}. None of the material presented in this section is new, and a reader familiar with this subject can jump directly to the next section, although it might be helpful in understanding the logic we will follow in the rest of the paper. 

We start with a short review of the Polchinski-Strassler (PS) mechanism, namely the polarization of D3 branes into D5 or NS5 branes in the supergravity dual of the mass-deformed $\mathcal{N}=4$ 4$d$ SYM. We will then consider the extension of this analysis to the mass-deformed 3d $\mathcal{N}=8$ theory on the volume of M2 branes, which is relevant for our discussion in the next section.

\subsection{The polarization of D3-branes in \texorpdfstring{$AdS_5 \times S^5$}{AdS5} (Polchinski-Strassler)}
\label{subsec:reviewPS}

We start by considering the low-energy world-volume theory of a stack of $N$ D3 branes. The $SO(6)$ isometry of the five-sphere in the dual $AdS_5 \times S^5$ geometry corresponds to the $R$-symmetry of gauge theory and rotates its six real scalars. In $\mathcal{N}=1$ language these scalars combine into complex scalar components of three chiral multiplets, $\Phi_{1,2,3}$. Each of the three chiral superfields has a fermion component, a Weyl spinor $\lambda_{i=1,2,3}$. Together with a fourth spinor, $\lambda_4$, the gaugino of the vector multiplet, they transform as a $\mathbf{4}$ under $SU(4)$, the ``fermionic version" of $SO(6)$. Giving arbitrary masses $m_{1,2,3}$ to the chiral superfields results in an $\mathcal{N}=1$ theory, while $\mathcal{N}=2$ requires $m_1=m_2$ and $m_3=0$. At the same time, adding a mass term $m^\prime$ for $\lambda_4$ necessarily breaks all of the supersymmetries, since this fermion field has no scalar superpartner. 

It was first noticed by Girardello, Petrini, Porrati and Zaffaroni (GPPZ) in \cite{Girardello:1999bd} that the mass deformation of the boundary theory, which corresponds to a three-form flux perturbation of the $AdS_5 \times S^5$ gravity dual, leads to a spacetime with an IR naked singularity, caused by the backreaction of this three-form on the metric.

It was argued later by Polchinski and Strassler in \cite{Polchinski:2000uf} that the singularity is cured by ``polarizing" via the Myers effect \cite{Myers:1999ps} the D3 branes into spherical 5-branes shells that stabilize themselves at a fixed position in $AdS_5$ ``shielding" effectively the IR singular region. This observation was confirmed in a probe limit calculation ignoring the D5's (or NS5's) backreaction on the geometry, which amounts to keeping the portion of the D3 brane charge carried by the D5's smaller compared to the total D3 charge, $n\ll N$. The ``polarization" potential consists of three terms with different powers of $r$, the radial distance ``orthogonal" to the D3's,\footnote{\label{r-rho} We use $r$ to denote a generic coordinate distance to the D3's or M2-branes. It will later be identified with a radial or angular coordinate on the $S^3$ (in the case of D3) or $S^4$ (in the case of M2) according to the different polarisation channels.} of the form
\begin{equation}
\label{V-D5NS5}
V_{\textrm {D5/NS5}}  = \hat a_2 \cdot n r^2 + \hat a_3 \cdot r^3 + \hat a_4 \cdot \frac{r^4}{n} + \ldots   \, ,
\end{equation}
where the labeling for the coefficients is chosen for later convenience, and the dots stand for subleading $\mathcal{O} \left(n^{-2} \right)$ terms, which can be neglected for $r \sim n$.

For sufficiently large D3 charge, $n$, these terms are detailed-balanced, and one can safely ignore other terms in the $1/n$ expansion. The large-$n$ condition amounts to $n^2 \gg g_s^2 N$ and thus does not contradict the $n\ll N$ requirement. The three terms have the following origin:
\begin{itemize}
\item The $n^{-1} \cdot r^4$ term, represents the mass difference between a stack of $n$ D3 branes dissolved in a 5-brane wrapped on an $S^2$ inside the $S^5$ and the same stack of D3 branes without the 5-brane. Since 3-brane and 5-brane masses add in quadratures, this term is always positive. 

\item The $r^3$ term comes from the $C_6$ term in the WZ action of the D5 brane (or the $B_6$ in the action of the NS5 brane). One can easily show that it is determined by the EOM of the RR 5-form in the $AdS_5 \times S^5$ background, which is equivalent to:
 \begin{equation}
 \label{PS-3form-EOM}
 {\rm d} \left( Z^{-1} \left( \star_6 G_3 - i G_3 \right) \right) = 0 \, ,
 \end{equation}  
 where $G_3$ is the complex 3-form flux (see footnote \ref{G_3}) and $Z$ is the warp function. Since the 3-from in \eqref{PS-3form-EOM} is both closed and co-closed, it can be fixed from its UV asymptotics, where 
 it is, in turn, uniquely determined by the three masses $m_1$,  $m_2$ and $m_3$. The 3-form then gives rise to the cubic term in the polarization potential, and is independent of the value of the warp factor. 
 
 \item The $n \cdot r^2$ term is the leading order term in the large $n$ expansion. It comes from the imperfect cancellation between the electric repulsion and the gravitational attraction that the $n$ D3 branes feel inside the perturbed $AdS_5$. In the supersymmetric PS solution this term can be computed by finding the superpotential that gives the $r^4$ and $r^3$ terms and observing that the full potential can be calculated from this superpotential. This term can also be evaluated by computing explicitly the backreaction of the three-forms on the metric, dilaton and five-form \cite{Freedman:2000xb}. When supersymmetry is broken, this term can receive two additional contributions, one from the gaugino mass and one from a traceless mass term for the scalar bilinears~\cite{Polchinski:2000uf,Freedman:2000xb,Zamora:2000ha}, which corresponds to an $L=2$ five-sphere harmonic in the bulk.

\end{itemize}

\subsection{The polarization of M2-branes in \texorpdfstring{$AdS_4 \times S^7$}{AdS4}}
\label{subsec:M2AdSpol}

We now review the perturbed $AdS_4 \times S^7$ solution dual to the mass-deformed $\mathcal{N}=8$ M2-brane theory originally studied in \cite{Bena:2000zb}. From the point of view of $\mathcal{N}=2$ supersymmetry in three dimensions, this theory has four hypermultiplets. Turning on four arbitrary masses for these hypermultiplets ($m_1$, $m_2$, $m_3$ and $m_4$) preserves four supercharges. When the masses are equal the supersymmetries get enhanced to 16. \cite{Pope:2003jp,Bena:2004jw,Lin:2004nb}. 

The 4-form flux perturbation dual to the hypermultiplet masses leads to a naked singularity in the IR~\cite{Pope:2003jp} which gets resolved by the polarization of the M2 branes into M5 branes~\cite{Bena:2000zb}.\footnote{The interpretation of the Myers effect from the perspective of the M2 brane theory was not available until the precise understanding of the field content carried out by~\cite{Aharony:2008ug}. The appearance of the fuzzy 3-spheres in the mass deformed theory was then consequently confirmed in~\cite{Gomis:2008vc}.} Unlike the $AdS_5$ example, this has been confirmed in~\cite{Bena:2004jw,Lin:2004nb} by finding a fully back-reacted solution. To calculate the polarization potential for the probe M5 brane with a non-zero M2 charge one may use either the M5 probe action of~\cite{Pasti:1997gx} in the (perturbed) M2 geometry (as done in \cite{Bena:2000zb}) or reduce both the solution and the probe to 10 dimensions along one of the M2 world-volume directions, and calculate the potential of the resulting D4 brane using the DBI action of the latter (as done in \cite{Klebanov:2010qs}). Both approaches yield the same polarization potential
\begin{equation}
\label{V-M5}
V_{\textrm M5}  = a_2 \cdot n r^2 + a_4 \cdot r^4 + a_6 \cdot \frac{r^6}{n} + \ldots   \, ,
\end{equation}
where the $\ldots$ stands for the subleading $\mathcal{O} \left(n^{-2} \right)$ terms. The origin of the terms in this potential is analogous to the origin of the terms in the PS potential. 
\begin{itemize}
\item The $n^{-1} \cdot r^6$ term is the difference between the mass of a stack of $n$ M2 branes dissolved into an M5 brane wrapping a three-sphere and the mass of the same stack without the M5 brane. Again, since the masses of M2 and M5 branes add in quadratures, this term is always positive.

\item The $r^4$  term can be traced to the expansion of the equation of motion for four-form field strength in a background sourced by M2 branes:
\begin{equation}
 \label{PS-4form-EOM}
 {\rm d} \left( Z_\textrm{M2}^{-1} \left( \star_8 F_4 - F_4 \right) \right) = 0 \, ,
 \end{equation} 
 where $Z_\textrm{M2}$ is the M2 warp factor and $F_4$ is the magnetic part of the 4-from flux. This equation is very similar to the equation the gives the $r^3$ term in the PS potential~\eqref{PS-3form-EOM}. 
 
If instead of a background sourced by M2 branes we had perturbed around a solution sourced by anti-M2 branes, this equation would become
\begin{equation}
 \label{PS-4form-EOM-anti}
 {\rm d} \left( Z_{\overline{\textrm{M2}}}^{-1} \left( \star_8 F_4 + F_4 \right) \right) = 0 \, ,
 \end{equation} 

Since the combination $ Z_{\overline{\textrm{M2}}}^{-1} \left( \star_8 F_4 + F_4 \right)$ is closed and co-closed, it only depends on the UV data, and is independent of the value of $Z_\textrm{M2}$.
 
\item The $n \cdot r^2$ term comes again from the imperfect cancellation of the gravitational attraction and the electric repulsion that the $n$ M2 branes feel in the perturbed geometry. In a supersymmetric solution this term can also be fixed by demanding that the full potential comes from a superpotential \cite{Bena:2000zb}. When supersymmetry is broken this term can receive an additional contribution from a traceless mass term for the scalar bilinears, which corresponds to an $L=2$ harmonic in the bulk.

\end{itemize}

Since all the terms in the polarization potential are independent of the location of the M2 branes that source the solution, this allows us to find the polarization potential of all the $N_2$ M2 branes that source the geometry to polarize into M5 branes by breaking them into $N_5$ bunches of $n$ M2 branes each, and treating each shell as a probe in the background sourced by the other shells. The full potential is therefore given by replacing $n$ in equation~(\ref{V-M5}) by $N_2/N_5$ and multiplying with an overall factor of $N_5$. The potential for all the M2 branes to polarize into a single M5 brane is then given by formally taking $N_5 = 1$, which, despite being out of the range of validity of the calculation, agrees with the formula driven from the fully-back-reacted solution \cite{Bena:2004jw,Lin:2004nb}. More details of this can be found in Section~V.B of~\cite{Polchinski:2000uf} and in Section~IV of  \cite{Bena:2000zb}.

This concludes our brief review of brane polarization in mass-deformed theories. We will now address the main problem of the paper, namely the study of the polarization of anti-M2 branes in the CGLP supersymmetric background.

\section{The polarization of anti-M2 branes in the CGLP geometry}
\label{subsec:antiM2cglp}

Let us now study the possible polarization of $N_{\overline{\textrm{M2}}}$ anti-M2 branes immersed in the CGLP background~\cite{Cvetic:2000db}  with $M$ units of self-dual flux, that we reviewed in Section~\ref{subsec:susyCGLP}.\footnote{In what follows we will always assume that $N_{\overline{\textrm{M2}}} \ll \widetilde M^2$, since otherwise the solution will not have positive M2 charge at infinity. Note that for anti-D3's in Klebanov-Strassler such an assumption is not necessary, because  the positive charge dissolved in the fluxes will always dominate asymptotically.}
Our goal is to study M5 brane polarization in a fully back-reacted anti-M2 geometry. 

For clarity, it is useful to review the various configurations that we will consider: as depicted in 
Figure~\ref{fig:Smeared-Localized}, the original un-polarized anti-M2 branes can be either  smeared over the non-vanishing $S^4$ at the tip of the CGLP solution (preserving therefore the symmetry of this solution), or can be fully-localized at a point on this $S^4$. With obvious Santa Claus bias, we will refer to this point as the North Pole. The only possible polarization channel of the smeared M2 branes is into M5 branes wrapping the  shrinking $S^3$ at finite distance away from the tip. We refer to this channel as the ``transverse channel''. When the branes are localized they can also polarize into M5 branes that wrap an $S^3$ inside the $S^4$ at the tip, and we refer to this as the Klebanov-Pufu (KP) channel.\footnote{Note that unlike (p,q) 5-branes which couple to a combination of $C_6$ and $B_6$, here there is a single type of coupling to $A_6$ and therefore these two channels are possibe.}  These notations are summarized in Table~\ref{table:smearloc}. 

\begin{table}
\centering
\begin{tabular}{ |r |c c c| c c c |c c c c |c |}\hline & \multicolumn{3}{|c|}{$\overbrace{\hspace{1.5cm}}^{\mathbb{R}^{1,2}}$} & \multicolumn{3}{|c|}{$\overbrace{\hspace{1.5cm}}^{S^3}$} & \multicolumn{4}{|c|}{$\overbrace{\hspace{2.2cm}}^{S^4}$} & $\overbrace{}^{\rho}$ \\ & 0 & 1 & 2 & 3 & 4 & 5 & 6 & 7 & 8 & 9 & 10 \\  \hline smeared M2 & $\times$ & $\times$ & $\times$ & $\cdot$ & $\cdot$ & $\cdot$ & $\sim$ & $\sim$ & $\sim$ & $\sim$ & $\cdot$\\ localized M2 & $\times$ & $\times$ & $\times$ & $\cdot$ & $\cdot$ & $\cdot$ & $\cdot$ & $\cdot$ & $\cdot$ & $\cdot$ & $\cdot$\\ transverse M5 & $\times$ & $\times$ & $\times$ & $\times$ & $\times$ & $\times$ & $\cdot$ & $\cdot$ & $\cdot$ & $\cdot$ & $\cdot$\\ Klebanov-Pufu M5 & $\times$ & $\times$ & $\times$ & $\cdot$ & $\cdot$ & $\cdot$ & $\times$ & $\times$ & $\times$ & $\cdot$ & $\cdot$\\ \hline \end{tabular}
\caption{Directions along which the branes extend ($\times$) or are smeared ($\sim$).} \label{table:smearloc}
\end{table}

Our strategy is to compute the polarization potential of the anti-M2 branes using the same logic as \cite{Polchinski:2000uf,Bena:2000zb}: we smear the anti-M2 branes and consider a region where the solution is of anti-M2 brane type. We then examine the perturbations of this region by transverse fluxes and metric modes that come from gluing it to the asymptotic region. We then calculate the potential for the smeared anti-M2 branes to polarize into M5 branes in the transverse channel. This section is devoted to the calculation of the polarization potential for this channel, while Section~\ref{sec:Localized} is devoted to the polarization potentials of localized anti-M2 branes.

\subsection{General approach}

Let us assume that one has already constructed a fully-back-reacted solution describing $N_{\overline{\textrm{M2}}}$  localized unpolarized anti-M2 branes in an asymptotically-CGLP geometry. Since near the North Pole the metric of the Stenzel space looks like $\mathbb{R}^8$, the backreaction of the anti-M2 branes should result in a ``small" $AdS_4 \times S^7$ throat with the radius fixed by $N_{\overline{\textrm{M2}}}$. However, this throat will not be a ``clean'' throat, since the gluing to the asymptotic CGLP solution will alter its UV region, and will introduce non-normalizable modes. In particular,  the self-dual (SD) flux of the CGLP solution will leak into the anti-M2 throat and try to polarize the anti-M2 branes into M5 branes, much as one expects from the probe computation in~\cite{Klebanov:2010qs}. It is very important to stress that this picture is also valid when the sources are smeared and the anti-M2 dominated region is no longer of the $AdS_4 \times S^7$ form.

Our strategy is to describe the physics of polarizing back-reacted anti-M2 branes using the Polchinski-Strassler method applied to M2 branes \cite{Bena:2000zb} that we reviewed in Section~\ref{subsec:M2AdSpol}. As we will show in the next subsections, we will recover this way precisely the same form and the same physical interpretation of the polarization potential as in~\eqref{V-M5}.

We begin by considering a region where the solution with unpolarized branes has anti-M2 character, in that its electric field and warp factor are such that a probe anti-M2 brane will feel (almost) no force. 
Since ASD flux is mutually supersymmetric with the anti-M2's, this field can in principle be arbitrarily large.\footnote{Although, as we will explain in Section \ref{sec:validity}, it turns out that its coefficients  will be much smaller than  \mbox{$\widetilde{M} \sim M l_\textrm{P}^{-3}$}.} Our strategy is to treat the 
SD flux coming from the gluing to the CGLP region as a perturbation on the anti-M2 solution, exactly as described in Section~\ref{subsec:M2AdSpol}.

When the anti-M2 branes are smeared on the Stenzel tip the full solution will have $SO(5)$ isometry, and will be ``anti-M2 dominated" between two constant-radial-coordinate hypersurfaces at $\rho_1$ and $\rho_2$. In general the fluxes that cause brane polarization become stronger in the infrared and (unless one takes brane polarization into account) give a naked singularity of GPPZ/Pope-Warner type \cite{Girardello:1999bd, Pope:2003jp}. The infrared hypersurface at $\rho=\rho_1$ is where the energy of these SD fluxes becomes stronger than that of the anti-M2 branes and the solution loses its anti-M2 character. Moreover, as we discussed in section~\eqref{subsubsec:ASD}, the ASD flux will also have a singular solution in the infrared, see (\ref{anti-SDflux}). We thus define $\rho_1$ such that for $\rho>\rho_1$ the energy density of the SD is small, and the one of the ASD flux is finite.

\begin{figure}[t]
\centering
\includegraphics[scale=0.5]{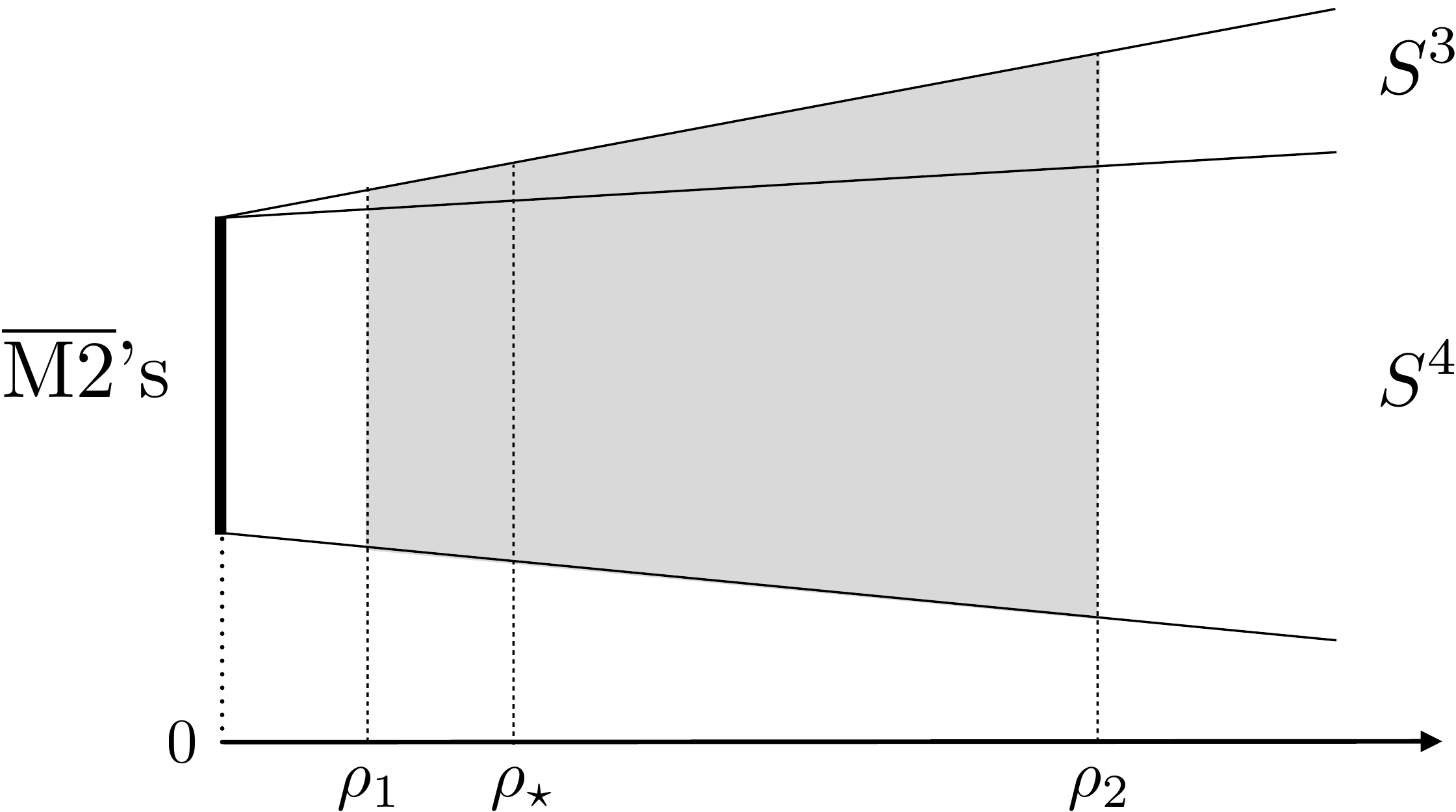}
\caption{In the interval $\rho_1 < \rho < \rho_2$ (marked in light grey) the energy of the SD flux is small compared to that of the electric flux of the anti-M2 branes and therefore can be treated perturbatively. The (UV) scale where this anti-M2 region is ``glued" to the (self-dual) CGLP solution is $\rho_2$, and the (IR) scale where the backreaction of the SD flux becomes important is $\rho_1$.
When the $\overline{\rm M2}$ sources are localized these slicings will be deformed and will have a non-trivial angular dependence.}
\label{fig:Setting}
\end{figure}

The ultraviolet hypersurface at  $\rho = \rho_2$ is where the anti-M2 dominated region is glued to the CGLP asymptotics. As we will discuss in the next section, when the anti-M2 branes are localized, the IR and UV boundaries of the brane dominated region will no longer be constant-$\rho$ hypersurfaces, but will get an angular dependence. For the sake of clarity, we postpone the evaluation of these scales and the discussion of the range of validity of our calculations to Section~\ref{sec:validity}.
Our first purpose is to calculate the polarization of a shell of M5 branes with anti-M2 charge dissolved in it at a radius $\rho_\star$ satisfying $\rho_1 \ll \rho_\star \ll \rho_2$, as depicted schematically in Figure \ref{fig:Setting}. To do this we will solve the equations of motion in this region order by order in an expansion in the SD flux parameter, $\mathcal{M}_\textrm{SD}$.

\subsection{The flux expansion} 

\label{sec:fluxexpansion}

We now apply the general strategy detailed above to compute the flux and the warp factor in the anti-M2-dominated region perturbed with SD flux. Remarkably, to compute the polarization potential in the transverse channel we will only need the leading-order expression of the modes $\xi_f^{-}$, $\xi_h^-$ and $\xi_z^-$. Since we consider smeared anti-M2's, the unperturbed solution, at zeroth-order in the flux expansion parameter $\mathcal{M}_\textrm{SD}$ is not  $AdS_4\times S^7$ but a warped geometry with ASD flux. 

To find this geometry one needs to solve $\xi^-_a=0$ for $a=z,f,h$, and it may appear that the only regular solution to these three equations is the warped Stenzel metric (\ref{Stenzel-functions}) with ASD flux  described in Equation (\ref{anti-SDflux}). However, this conclusion is a bit too hasty: the vanishing of $\xi_z^-$, $\xi_f^-$ and $\xi_h^-$ does not imply that the remaining $\xi$-functions, $\xi_\alpha^-$, $\xi_\beta^-$ and $\xi_\gamma^-$, are zero as well. The second-order equations in (\ref{xi-abc-dot}) have six integration constants. We have already explained in Section~\ref{sec:Setup} how to fix the three constants of the SUSY solution for which $\xi^-_\alpha,\xi^-_\beta,\xi^-_\gamma=0$. Intuitively one can use the other three constants to construct a new solution that smoothly approaches the Stenzel metric functions (\ref{Stenzel-metric}) both in the UV and in the IR. In fact, this is equivalent to finding a new Ricci flat metric within the Ansatz (\ref{Stenzel-metric}). Abusing notation, we will continue to refer to this new metric as ``Stenzel'' and to the large $S^4$ at its tip as the ``Stenzel tip''. This new metric does not have to be K\"{a}hler\footnote{A similar calculation for the conifold has been done in \cite{Dymarsky:2011ve,Bennett:2011va}.} and for our purposes we will not need to know its exact form,\footnote{The proof of this fact is identical to the (slightly more complicated) proof we presented in Section~\ref{sec:interpolate} (see the paragraph below (\ref{xi-zfhRegIR})), and we will not repeat it here.} but just the leading-order expansion of the function $e^{\beta-\alpha} = \rho+ \ldots$. 

We now have to find the zeroth-order solution for the warp function. It follows from the $\xi^-_z=0$ equation in (\ref{xi-zfh}) that: 
\begin{equation}
e^{3 z_0} = 6 \int_{\rho}^{\infty} \d \bar{\rho} \, e^{-3 ( \alpha_0 + \beta_0)} \left(h_\textrm{ASD} \left( f_\textrm{ASD} - 2 h_\textrm{ASD} \right)-P \right) \, ,
\end{equation}
where we use ${}_0$ subscripts to distinguish between the Stenzel solution (\ref{Stenzel-functions}) and our zeroth-order functions, and the parameter $P$ is proportional to the number of anti-M2 branes, $N_{\overline{\textrm{M2}}}$.
One can readily check that for zero $h$ and $f$,  $e^{3 z_0} \sim   \rho^{-2} + \ldots$, as expected. As we have explained in the previous section, in general, the ASD flux does not have to vanish in the brane-dominated region. Therefore, for sufficiently small $\rho$ the contribution of the IR singular ASD flux will alter this behavior of the warp function. This corresponds to the $\rho \ll \rho_1$ region on Figure \ref{fig:Setting}.

We now perturb this solution with SD flux. For this we will need to turn on the modes $\xi_a^{-}$, which are strictly zero for a solution with only ASD flux. At first order in $\mathcal{M}_\textrm{SD}$ we only have to solve the equations for ${\xi_f^-}^\prime$ and ${\xi_h^-}^\prime$ in  (\ref{xi-fhz-dot}), since the metric functions and $\xi_z^-$ remain the same. Note that this is equivalent to solving~\eqref{PS-4form-EOM-anti} in the brane dominated region. These equations reduce to:
\begin{equation}
\label{sol-xif-xih-Exact}
 {\xi_f^-}^\prime = - \frac{1}{2} \coth \rho \,  \xi_h^-
\, , \qquad \qquad
 {\xi_h^-}^\prime = - 6 \tanh^3 \rho \, \xi_f^- + 2 \coth \rho \, \xi_h^-  \, ,
\end{equation}
where we have explicitly used the Stenzel value of $e^{\beta-\alpha}=\tanh(\rho)$, although, as we mentioned earlier, only  $e^{\beta-\alpha}=\rho+\ldots$ is relevant for the conclusions. Note that we have already encountered these equations in Section~\ref{sec:interpolate}.

The solution of (\ref{sol-xif-xih-Exact}) is:
\begin{eqnarray}
\label{sol-xif-xih}
\frac{\xi_f^-}{\mathcal{M}_\textrm{SD}} &=& b_1 \cosh^3 \rho + b_2 \dfrac{\cosh^4 \rho - 1}{\cosh \rho}
\, , \qquad \\
\frac{\xi_h^-}{\mathcal{M}_\textrm{SD}} &=& -6 b_1 \sinh^2 \rho \cosh \rho - 2  b_2 \dfrac{\sinh^2 \rho \left(3 \cosh^4 \rho + 1 \right) }{\cosh^3 \rho} \nn \, ,
\end{eqnarray}
where we have explicitly factored out $\mathcal{M}_\textrm{SD}$  in order to make $b_1$ and $b_2$ independent of the expansion parameter. Notice that from the definition of $\xi_f^-$ and $\xi_h^-$ in (\ref{xi-zfh}) it is clear that the $\mathcal{M}_\textrm{SD}$ parameter corresponds to the SD part of the magnetic flux divided by the number of the anti-M2's.

In the next section we will need only the small-$\rho$ expansion of this solution, which is:
\begin{equation}
\label{sol-xif-xih-Expansion}
\frac{\xi_f^-}{\mathcal{M}_\textrm{SD}} = b_1 + \mathcal{O} \left(\rho^2 \right)
\, , \qquad 
\frac{\xi_h^-}{\mathcal{M}_\textrm{SD}} = \left( -6 b_1 - 8 b_2 \right) \rho^2 + \mathcal{O} \left(\rho^4 \right) \,.
\end{equation}

We now look for the solution at order $\mathcal{M}_\textrm{SD}^2$. One of the remarkable results of the next section is that to compute the polarization potential for M5 branes we will only need to know $\xi_z^{-}$ at second order in $\mathcal{M}_\textrm{SD}$, and no other metric $\xi^-$'s. Hence, the only equation we need to solve at order $\mathcal{M}_\textrm{SD}^2$ is the last equation in (\ref{xi-fhz-dot}).

Finding an explicit expression for ${\xi_z^-}^\prime$ is unlikely since the warp function $e^{3 z_0}$ is itself written in terms of an unsolvable integral. We, however, only need a leading term in the Taylor expansion for small $\rho$. To find this term we first have to notice that the $K^\prime e^{3z}$ factor in the $\xi_z^{-}$ equation multiplies $\xi_z^{-}$ which is already of the second order, and so we can use the zeroth-order equation $\xi_z^-=0$ to replace $K_0^\prime e^{3z_0}$ by $-3 z_0^\prime$. The ${\xi_z^-}^\prime$ equation can then by rewritten as:
\begin{equation}
\left( e^{-3 z_0} \xi_z^- \right)^\prime = - \frac{1}{4} \left( 12 e^{3 (\beta_0-\alpha_0)} {\xi_f^-}^2  +  e^{\alpha_0-\beta_0} {\xi_h^-}^2 \right) \, .
\end{equation}
This leads to the final result:
\begin{equation}
\label{Msd-SecondOrder}
\frac{e^{-3 z_0} \xi_z^-}{\mathcal{M}_\textrm{SD}^2} = -  \left( 3 b_1^2 + 6 b_1 b_2 + 4 b_2^2 \right) \rho^4 + \ldots  \, .
\end{equation}
Here we omitted a homogeneous part of the solution proportional to $e^{3 z_0}$, since it corresponds to a non-physical singularity.
 
We will use this expression to estimate the polarization potential in the upcoming section.

\subsection{The polarization potential}
\label{subsec:PolarizationPotential}

In this section we use the result obtained above to compute the polarization potential for all the smeared anti-M2 branes to polarize into M5 branes wrapping the shrinking $S^3$ of the Stenzel geometry at a finite radius (see Table \ref{table:smearloc}). As in  \cite{Polchinski:2000uf,Bena:2000zb}, we will first compute the action of a probe M5 brane with  anti-M2 charge $n$ in the back-reacted throat geometry sourced by the rest of the anti-M2 branes and described in detail in Section~\ref{sec:generalapproach}. We will then argue that the potential for this probe M5 brane is independent of the location of the anti-M2 branes that source the solution, and hence this potential give the fully-back-reacted polarization potential in the transverse channel, both for smeared and for localized anti-M2 branes. Finally, in Section~\ref{sec:Localized} we will use this result to infer the M5 potential in the Klebanov-Pufu channel in the geometry of localized sources.

There are two ways of computing the potential of a probe M5 with anti-M2 charge in a certain eleven-dimensional supergravity background. The first is to use directly the M5 action of Pasti, Sorokin and Tonin, \cite{Pasti:1997gx}, as was done, for instance, in \cite{Bena:2000zb}. The second is to reduce both the background and the probe to type IIA string theory, compute the potential of the resulting probe (a D4 brane with F1 charge dissolved in it) in the resulting background, and then reinterpret this as the potential for the M2-M5 polarization. This was done for example in~\cite{Klebanov:2010qs}.
The two approaches give the same answer, but given the relative complexity of the Pasti-Sorokin-Tonin action, we find it more instructive to compute the potential using the second approach.

\subsubsection{The IIA reduction of the 11-dimensional background}

Our strategy is to reduce both the background of Section~\ref{sec:Setup} and the M5-M2 probe to type II string theory along one of the M2 world-volume coordinates, say $x^2$. The M5-M2 probe becomes a D4 brane wrapping the shrinking Stenzel $S^3$ with $n$ anti-F1 strings dissolved in it. Reducing the background of Section~\ref{sec:Setup} gives the $10d$ metric, dilaton and $B$-field:
\begin{equation}
\d s_{10}^2 = e^{-3 z} \d x_\mu \d x^\mu + \d s_8^2 \, ,
\qquad
e^\phi = e^{-\frac{3}{2} z} \, ,
\qquad
B_{2} = K \d x_0 \wedge\d x_1 \, ,
\end{equation}
where the eight-dimensional Stenzel metric was given in (\ref{Stenzel-metric}) and the function $K(\rho)$ appears in (\ref{K-prime}). In order to compute the polarization potential we only need the IIA R-R four-form field strength with legs on the shrinking $S^3$:  $F_4 = f^\prime d\rho\wedge \tilde{\sigma}_1 \wedge \tilde{\sigma}_2 \wedge \tilde{\sigma}_3 + \ldots$. The other components (denoted by $\ldots$) can be computed straightforwardly from (\ref{4-form}).
The forms that enter in the polarization potential are given by:
\begin{eqnarray}
C_3 &=& f \, \tilde{\sigma}_1 \wedge \tilde{\sigma}_2 \wedge \tilde{\sigma}_3 + \ldots 
\nonumber \\
\d \left( C_5 + B_2 \wedge C_3 \right) &=& \, \d x_0  \wedge \d x_1  \wedge \left( K F_4 + e^{-3 z} \star_8 F_4 \right) =
\\
&=& \, \d x_0  \wedge \d x_1  \wedge \left[ \left( K {f^\prime} - 6 e^{-3 (\alpha-\beta+z)} h \right) \d \rho \wedge \tilde{\sigma}_1 \wedge \tilde{\sigma}_2 \wedge \tilde{\sigma}_3 + \ldots \right] \, .
\nonumber 
\end{eqnarray}

\subsubsection{The probe action}

In order to compute the potential governing the polarization of anti-F1 strings into a D4 brane wrapping the shrinking Stenzel 3-cycle, we start from the action describing a D4 brane with a non-trivial world-volume electric field:
\begin{equation}
S_\textrm{D4} = \int_{\mathbb{R}^{1,1}}  \mathcal{L}_\textrm{D4} = \mu_4 \int_{\mathbb{R}^{1,1} \times S^3} \left[ - e^{-\phi} \sqrt{-\textrm{det} \left(g_{ab} + 2 \pi l_s^2 \mathcal{F}_{ab} \right)} +  \left( C_5 + 2 \pi l_s^2 \mathcal{F}_2 \wedge C_3 \right) \right] \, ,
\end{equation}
where $ 2 \pi l_s^2 \mathcal{F}_2 =  2 \pi l_s^2 F_2 + B_2$ and $F_2$ is the electric field:
\begin{equation}
2 \pi l_s^2 F_2 = \mathcal{E} \, \d x_0 \wedge\d x_1 \, .
\end{equation}
Plugging in the metric, the dilaton and the forms computed in the previous subsection, and integrating over the 3-sphere we find the Lagrangian density:
\begin{eqnarray} \label{LD4}
\mathcal{L}_\textrm{D4} (\mathcal{E}) &=& \mu_4 V_{S^3}  \Bigg[ - e^{3 \beta + \frac{3}{2} z} \sqrt{e^{-6 z}- (\mathcal{E}+K)^2} + \mathcal{E} f  
\nonumber \\
&&
\qquad \qquad \qquad \qquad
+ \int_{0}^{\rho} d\bar\rho\left( 2 e^{3 (\beta-\alpha)} \xi_f^- +  (K - e^{-3 z}) {f^\prime} \right) \Bigg] \, . 
\end{eqnarray}
Here $V_{S^3}$ stands for the 3-sphere volume, $V_{S^3}=\int \widetilde{\sigma_1} \wedge \widetilde{\sigma_2} \wedge \widetilde{\sigma_3}$.
In deriving this formula we performed an integration by parts and used the definition of $\xi_f^-$ in (\ref{xi-zfh}).

The fundamental string charge of the D4 brane (which corresponds in eleven dimensions to the M2 charge of the M5 brane) is the momentum conjugate of the world-volume electric field:
\begin{equation} 
\label{LT}
n \equiv - \frac{\partial \mathcal{L}_\textrm{D4} (\mathcal{E})}{\mu_\textrm{F1} \partial \mathcal{E}} \, ,
\end{equation}
where the minus sign is introduced for later convenience, and $\mu_\textrm{F1}$ appears in the definition because the fundamental string coupling to the $B$-field is given by $\mu_\textrm{F1} \int B$.

To compute the potential of this D4 brane we need to find the Hamiltonian corresponding to this action, and to do this we begin by expressing $\mathcal{E}$ in terms of $n$ using (\ref{LT}):
\begin{equation}
\mathcal{E}+K = e^{-3 z} \left( 1 + \dfrac{ \left( {l_\textrm{P}^{-3} V_{S^3}} \right)^2 e^{6\beta + 3 z}}{\left(n + l_\textrm{P}^{-3} V_{S^3} \, f\right)^2} \right)^{-1/2} \, .
\end{equation}
The Hamiltonian is then given by the Legendre transform:
\begin{eqnarray}
\label{L_n}
\mathcal{ H}_\textrm{D4}(n) = -n \mathcal{E} - \mathcal{L}_\textrm{D4} (\mathcal{E})
 &=& - \left(n + l_\textrm{P}^{-3} V_{S^3} \, f\right) \left[ e^{-3 z} \left( 1 + \dfrac{ \left( {l_\textrm{P}^{-3} V_{S^3}} \right)^2 e^{6\beta + 3 z}}{\left(n + l_\textrm{P}^{-3} V_{S^3}  \, f\right)^2} \right)^{1/2} - K \right] 
 \nonumber \\
 && \quad
 - l_\textrm{P}^{-3} V_{S^3} \int^{\rho}_{0} \d \bar{\rho} \left( 2 e^{3(\beta-\alpha)} \xi_f^- +  \left(K - e^{-3z} \right) {f^\prime} \right) \, .
\end{eqnarray} 

Note that in deriving this expression we have used the fact that $\mu_4/\mu_\textrm{F1}=l_\textrm{P}^{-3}$. The fact that resulting Hamiltonian depends only on the eleven-dimensional Planck scale, $l_\textrm{P}$, and is independent of the compactification radius $l_{11}$, confirms the validity of this approach to compute the M2-M5 polarization potential. Note also  that in our conventions the flux functions $f$ and $h$ have dimension $(\textrm{length})^3$ as evident from (\ref{P}) and (\ref{K-prime}). This is different from the conventions of most of the literature, where $f$ and $h$ are dimensionless (see the remark below (\ref{4-form})).

Up to order $\mathcal{M}_\textrm{SD}^2$ and $n^{-1}$, the polarization potential, $V=-\mathcal{H}(n)$, is:
\begin{equation}
\label{Potential-not-final}
V = \left( e^{-3z} - K \right) \cdot n + 2 l_\textrm{P}^{-3} V_{S^3} \int^\rho_0 e^{3 (\beta_0 - \alpha_0)} \xi_f^-  \d \bar{\rho} + \dfrac{1}{2} \left( l_\textrm{P}^{-3} V_{S^3} \right)^2 e^{6 \beta_0} \cdot \dfrac{1}{n} + \mathcal{O}\left( n^{-2}, \mathcal{M}_\textrm{SD}^3\right) \, ,
\end{equation}
where the ${}_0$ index denotes the zeroth-order (no SD flux) solution. In deriving this result we omitted the $\left(K - e^{-3z} \right) f^\prime$ in the second line of (\ref{L_n}). We explain the reason at the very end of this subsection.

Keeping only the lowest terms in the $\rho$ Taylor expansion of all the functions, the potential has the expected M2-M5 form (\ref{V-M5}):
\begin{equation}
\label{Potential-final}
V = n a_2 \rho^2 + a_4 \rho^4 + \dfrac{a_6}{n} \rho^6 \, ,
\end{equation}
where:
\begin{eqnarray}
a_2 &=& - \dfrac{2}{3} \lim_{\rho \to 0}  \left( \dfrac{1}{\rho^2}  \int^\rho_0 e^{-3 (\alpha_0 + \beta_0 + z_0)} \xi_z^- \d \bar{\rho} \right) \, ,
\nonumber \\
a_4 &=& 2 \left(l_\textrm{P}^{-3} V_{S^3}\right) \lim_{\rho \to 0}  \left( \dfrac{1}{\rho^4}  \int^\rho_0 e^{3 ( \beta_0 - \alpha_0)} \xi_f^- \d \bar{\rho} \right) \, ,
 \\
a_6 &=&  \dfrac{1}{2} \left(l_\textrm{P}^{-3} V_{S^3}\right)^2 \lim_{\rho \to 0}  \left( \dfrac{1}{\rho^6}  \int^\rho_0 e^{6 \beta_0} \d \bar{\rho} \right) \, .
\nonumber
\end{eqnarray} 
Using the explicit results of the previous subsection for $\xi_f^-$ and $\xi_z^-$, we can now express these constants in terms of the parameters $b_1$ and $b_2$ introduced in (\ref{sol-xif-xih-Expansion}):
\begin{eqnarray}
\label{c2c4c6}
a_2 &=& \dfrac{\mathcal{M}_\textrm{SD}^2}{3} e^{-6\alpha_0(0)} \cdot \left( 3 b_1^2 + 6 b_1 b_2 + 4 b_2^2 \right) \, ,
\nonumber \\
a_4 &=& \dfrac{\mathcal{M}_\textrm{SD}}{2} l_\textrm{P}^{-3} V_{S^3} \cdot b_1 \, ,
\\
a_6 &=& \dfrac{1}{2} \left( l_\textrm{P}^{-3} V_{S^3} \right)^2 e^{6 \alpha_0(0)}    \, .
\nonumber
\end{eqnarray}

The first thing to observe about this potential is that its terms are detailed-balanced: at the radius where any two of its terms are equal, the remaining term is also of the same order. Indeed, the coefficients $b_1$ and $b_2$ are by construction $\mathcal{M}_\textrm{SD}$ independent, and it is easy to see that the geometric mean of the first and the third term is always of the order of the second one. It is also easy to see that at the detailed-balance scale, 
\begin{equation}
\label{rhostar}
\rho_\star^2 \sim n \mathcal{M}_{\textrm{SD}} l_\textrm{P}^{3} \, ,
\end{equation}
all the terms of order $\mathcal{O}(n^{-2})$ and/or $\mathcal{O}(\mathcal{M}_\textrm{SD}^3)$ and higher that we ignored can be safely ignored.

It is also straightforward to verify that the potential (\ref{Potential-final}) has {\it no (local) minimum} away from $\rho=0$. The condition for having the minimum is $a_4 < 0$ and $a_4^2 - 3 a_2 a_6 > 0$. While the former might be achieved by $b_1<0$, the second is equivalent to $\frac{5}{4} b_1^2 + 3 b_1 b_2 + 2 b_2^2 < 0$ which does not hold for any real $b_1$ and $b_2$.  We thus conclude that one of the possible polarization channels, the transverse channel, is absent for smeared anti-M2 branes.

Before closing the section, let us explain why in going from (\ref{L_n}) to (\ref{Potential-not-final}) we have ignored the $\left(K - e^{-3z} \right) f^\prime$ term in the potential. We know for example from Equation (\ref{Msd-SecondOrder}) that $\xi_z^-$ receives corrections at second order in the $\mathcal{M}_\textrm{SD}$ expansion. From the definition of $\xi_z^-$ in~(\ref{xi-zfh}), one can see that $\left(K - e^{-3z} \right)$ is also of order $\mathcal{M}_\textrm{SD}^2$. Nevertheless, this does not imply that this term is of the same order as the $\rho^2$ term in (\ref{Potential-final}). Indeed, a closer look at the Taylor expansion of $f^\prime$ reveals that the lowest term comes from its ASD part (\ref{anti-SDflux}), $f^\prime \approx 3 \left( \widetilde{C}_1 + \widetilde{C}_2 \right) \rho + \ldots$ and hence the $\left(K - e^{-3z} \right) f^\prime$ contribution to (\ref{Potential-final}) starts with a term of order $\rho^4$. Since the $\rho^4$ term in (\ref{Potential-final}) is by construction of order $\mathcal{M}_\textrm{SD}$, the $\left(K - e^{-3z} \right) f^\prime$ contribution is indeed negligible if  $\widetilde{C}_1 + \widetilde{C}_2$ is of order one or lower. In Section~\ref{sec:validity} we show that we work indeed in this regime. 

\section{Localized versus smeared sources}
\label{sec:Localized}

Having computed the smeared M2 brane polarization potential in the transverse channel, we will now try to use this calculation to learn about the polarization of anti-M2 branes localized at the North Pole on the 4-sphere (see Figure \ref{fig:Smeared-Localized}). As we discussed in Section~\ref{subsec:antiM2cglp} and as one can see from Table~\ref{table:smearloc}, these branes have two polarization channels: the transverse one, corresponding to M5 branes  wrapping the shrinking $S^3$, and the  Klebanov-Pufu channel, corresponding to M5 branes wrapping an $S^3$ inside the 4-sphere.

\subsection{The transverse channel}

To proceed, it is worth recalling one of the main results of~\cite{Polchinski:2000uf}: the polarization potential (\ref{V-D5NS5}) is independent of the actual value of the D3-brane warp factor, and hence it is the same regardless of the positions of the D3 branes that source the geometry. The proof of this statement involves a few steps. First, one notices that the closed (and co-closed) 3-from $Z^{-1} \left( \star_6 G_3 - i G_3 \right) $ in (\ref{PS-3form-EOM}) is fixed uniquely by its asymptotic UV value and is therefore independent of the warp factor $Z$, which encodes the information about the source distribution. This guarantees that the $r^3$ term in the potential is $Z$-independent. Next, one argues that the same observation holds for the $r^4$ term, which measures the 5-brane mass increment and is proportional to the square of the volume (in un-warped coordinates) of the sphere on which the 5-brane is wrapped.\footnote{The extra factors of $Z$ cancel each other leaving a warp-function independent contribution, see (62) of~\cite{Polchinski:2000uf} and (29a) of~\cite{Bena:2000zb}.}  Finally, the $r^2$ term is also $Z$-independent - this can be seen either by invoking supersymmetry, or by realizing that this term comes from non-normalizable $AdS$ modes corresponding to boson masses. 

One can make exactly the same argument about the polarization potential of anti-M2 branes into M5 branes. As explained in Section \ref{subsec:M2AdSpol} the $r^4$ term comes from the self-dual four-form $Z^{-1} \left( \star_8 F_4 + F_4 \right) $, which is closed and co-closed (\ref{PS-4form-EOM}) and therefore independent of the value of the warp factor $Z$. The $r^6$ term is proportional to the square of the volume of the sphere that the M5 branes wrap, and is again $Z$-independent. Finally, the $r^2$ term encodes the boson masses, and it also $Z$-independent. These results have far-reaching consequences for understanding brane polarization, as they indicate that one can use the probe calculation to find the polarization potential of all the M2 branes that make up the solution, by splitting them in several bunches $n$ M2 branes and finding the polarization action of each bunch in the background sourced by the other bunches. 

One can now apply these results to argue that the polarization potential of localized CGLP anti-M2-branes in the transverse channel is the same as that of smeared anti-M2 branes: To do this one should first remember that the asymptotic value of $Z^{-1} \left( \star_8 F_4 + F_4 \right)$, and hence the $r^4$ term in the potential, is uniquely determined by the fermion masses in the theory on the two-branes \cite{Bena:2000zb}. In our solution, these four masses are determined by the gluing of the anti-M2-dominated region to the CGLP geometry at the hypersurface at $\rho=\rho_2$. Similarly, the $r^2$ terms in the potential corresponds to boson masses in the  two-brane theory and are also determined by the gluing at this hypersurface.

As we mentioned before, when the anti-branes are localized, this hypersurface will not be at constant $\rho$ any more, but rather will acquire some non-trivial angular dependence. In general, one expects this to affect the (co)closed self-dual 4-form (\ref{PS-4form-EOM-anti}) and the metric, and hence to modify the $\rho^4$ and $\rho^2$ terms in the potential. However, as we will discuss in detail in Section~\ref{sec:validity}, this modification becomes negligible when the gluing scale $\rho_2$ is much larger than the size of the blown-up Stenzel 4-sphere, and this can be easily achieved by properly tuning the free parameters of the solution: $N_{\overline{\rm M2}}$, $\widetilde{M}$ and the 4-sphere radius. In this limit the transverse-channel polarization potential for localized anti-M2 branes is exactly that of smeared sources, given in (\ref{Potential-final})-(\ref{c2c4c6}), which has no minimum at a finite $\rho$.\footnote{On the other hand, we are assuming that polarisation, if it happens at all, occurs in the region $\rho>\rho_1$. Otherwise this mechanism would not cure the singularity, and the anti-M2 solution is clearly unphysical.}  From now on we will assume that we work in this limit. 

\subsection{The Klebanov-Pufu channel}

Having obtained the polarization potential of localized back-reacted branes in the transverse channel, we can now use the physics of brane polarization to obtain the back-reacted polarization potential in the Klebanov-Pufu channel~\cite{Klebanov:2010qs}.

To do this we first need to use some convenient coordinates near the localized sources. The Stenzel space is defined by:
\begin{equation}
\sum_{i=1}^{5} z_i^2 = \epsilon^2 \, .
\end{equation}
In terms of $x_i \equiv \textrm{Re} (z_i)$ and $y_i \equiv \textrm{Im} (z_i)$ it translates into:
\begin{equation}
\sum_{i=1}^{5} x_i^2 - \sum_{i=1}^{5} y_i^2 = \epsilon^2 
\qquad \textrm{and} \qquad
\sum_{i=1}^{5} x_i  y_i = 0 \, .
\end{equation}
At the North Pole of the 4-sphere we have $x_1=\epsilon$ and $y_1=0$, while the remaining eight parameters $\left( x_2, \ldots, x_5, y_2, \ldots, y_5,\right)$ provide a good set of $\mathbb{R}^8$ coordinates in the vicinity of the pole. These branes break the isometry group of Stenzel space from $SO(5)$ down to an $SO(4)$ which simultaneously rotates $\left( x_2, \ldots, x_5 \right)$ and $\left( y_2, \ldots, y_5 \right)$. There are three invariants of  this rotation group:
\begin{equation}
\label{ThreeComb}
 \theta^2 \equiv \sum_{i=2}^{5} x_i^2\, ,  \qquad
\sum_{i=2}^{5} x_i y_i\,  \qquad {\rm and} \qquad \rho^2 = \sum_{i=2}^{5} y_i^2\, ,
\end{equation}  
the last one being the Stenzel radial coordinate we are used to. 

Our strategy is to first consider certain polarization channels where the scale of brane polarization is smaller than the size of the four-sphere\footnote{As we will explain in Section \ref{sec:validity}, this can be easily achieved by increasing the M5 dipole charge of the polarizing shell.} and therefore the anti-M2 branes at the North Pole can be treated as anti-M2 branes in $\mathbb{R}^8$. Since we are in a region where the anti-M2 branes dominate the geometry, the region where polarization will happen will be a small $AdS_4\times S^7$ around the CGLP tip North Pole, and we can therefore use all the techniques for studying M2-brane polarization discussed in section \ref{subsec:M2AdSpol} to compute the polarization potential for all channels.

In particular, we know that when the four fermion masses of the M2 brane theory are equal, the M2 branes polarize into M5 branes wrapping three-spheres. Hence, the $SO(4)$ symmetry of the solution implies that this small $AdS_4\times S^7$ is perturbed with equal fermion masses. By itself this perturbation would be supersymmetric, and give rise to a potential that is a perfect square, both for the transverse channel and for the KP channel. However, as we discussed in Section ~\ref{subsec:PolarizationPotential}, the polarization potential receives also a non-supersymmetric contribution from a traceless boson mass bilinear, which corresponds to an $L=2$ term on the $S^7$.
Hence, the generic polarization potential in the transverse channel can be written as
\begin{equation}
\label{V-transverse}
V^{\rm T}(\rho) = V^{\rm T}_{\rm SUSY}(\rho) + V^{\rm T}_{L=2}(\rho)\, ,
\end{equation}
while the potential in the Klabanov-Pufu channel can be written as
\begin{equation}
\label{V-KP}
V^{\rm KP}(\theta) = V^{\rm KP}_{\rm SUSY}(\theta) + V^{\rm KP}_{L=2}(\theta)\, .
\end{equation}
As discussed above and as shown explicitly in \cite{Bena:2000zb,Bena:2004jw,Lin:2004nb}, the supersymmetric polarization potential is the same in the two channels:
\begin{equation}
\label{V-equal-susy}
V^{\rm KP}_{\rm SUSY}(x) =V^{\rm T}_{\rm SUSY}(x) .
\end{equation}

The story is a bit more subtle for the non-supersymmetric $L=2$ contribution to the potential. The $SO(4)$ isometry of the configuration constrains it to be a combination of only two harmonics. Indeed, (\ref{ThreeComb}) lists all possible $SO(4)$ invariant quadratic combinations of $x_i$'s and $y_i$'s,
while the traceless requirement further implies that the coefficients of $\rho^2$ and $\theta^2$ sum up to zero:    
\begin{equation}
\label{General-Pol-Pot}
V_{L=2} = \mu_\textrm{diag} \cdot \left(\rho^2 -  \theta^2  \right)
+ \mu_\textrm{off-diag} \cdot \sum_{i=2}^{5} x_i y_i \, .
\end{equation}
Since for the transverse channel we have $x_{2,3,4,5}=0$, while for the KP channel $y_{2,3,4,5}=0$, we can see that the off-diagonal contribution drops off, and therefore the two channels will receive equal and opposite contributions:
\begin{equation}
V^{\rm KP}_{L=2}(\rho) = \mu_\textrm{diag} \rho^2 \, , \qquad V^{\rm KP}_{L=2}(\theta) = - \mu_\textrm{diag} \theta^2 \, .
\end{equation}

We are now able to extract the polarization potential for the KP channel from (\ref{Potential-final}), by writing 
\begin{equation}
V^T(\rho)= a_2 \rho^2 + a_4 \rho^4 + a_6 \rho^6 = \rho^2 \left( \sqrt{a_6} \rho^2 + \dfrac{a_4}{2 \sqrt{a_6}}\right)^2 - \left( \dfrac{a_4^2}{4 a_6} -a_2 \right) \rho^2 \, ,
\end{equation}
which combined with (\ref{V-transverse}),(\ref{V-KP}),(\ref{V-equal-susy}) gives
\begin{equation}
V^{\rm KP}(\theta) =  \theta^2 \left( \sqrt{a_6} \theta^2 + \dfrac{a_4}{2 \sqrt{a_6}}\right)^2 + \left( \dfrac{a_4^2}{4 a_6} -a_2 \right) \theta^2 \equiv  a_6 \theta^6  + a_4 \theta^4 +  \widetilde a_2 \theta^2\, .
\label{locpotential}
\end{equation}

The coefficient of the $\theta^2$ term, which cannot be captured in the probe approximation used in \cite{Klebanov:2010qs}, is therefore 
\begin{equation}
\label{c2c4c6-Tilde}
\widetilde{a}_2 = \dfrac{a_4^2}{2 a_6} -a_2 = - \dfrac{3}{4} e^{-6\alpha_0(0)} \cdot \left( b_1 + \dfrac{4}{3} b_2 \right)^2 
\, .
\end{equation}
This is a most striking result: for any value of $b_1$ and $b_2$, and therefore irrespective of the way the anti-M2 region is glued to CGLP, the quadratic term in the potential that describes the polarization of anti-M2 branes into M5 branes wrapping the $S^3$ inside the Stenzel $S^4$ tip is never positive. As we discussed in Section \ref{sec:generalapproach}, this term also gives the force felt by a probe anti-M2 brane in the background sourced by a stack of anti-M2 branes localized on the Stenzel tip. Equation (\ref{c2c4c6-Tilde}) implies that this force is always \emph{repulsive}\footnote{To be pedantic, there is of course a measure-zero possibility that $ b_1 = -\frac{4}{3} b_2$ and hence this force is zero. However, it is hard to see why such a miraculous cancelation will happen in a non-supersymmetric solution.} and hence anti-M2 branes at the bottom of the CGLP solution are \emph{tachyonic}! This is the main result of our paper.

\section{Range of validity}
\label{sec:validity}

In this section we discuss the approximations we have used in getting to our result, and its range of validity. In the absence of an explicit fully back-reacted solution that has CGLP asymptotics and anti-M2 branes in the infrared, we have used the fact that there should exist a region where the physics is dominated by the anti-M2 branes. There are several ways to try to define such a region, perhaps the most precise one is to require that the energy of the self-dual (SD) magnetic four-form fluxes be smaller than that of the electric four-form sourced by the anti-M2 branes. This allows one in turn to treat the fluxes as a perturbation around a BPS anti-M2 solution, and to argue that they satisfy equation 
(\ref{PS-4form-EOM-anti}), which is the key formula that allows one to compute the polarization potential of the fully-back-reacted branes. Another way to think about this region is as the region where a probe anti-M2 brane will feel (almost) no force because of the gravitational-electromagnetic cancellation in its action. Note that in this region the total magnetic $F_4$ flux is not necessarily small: this flux can have both a self-dual and an anti-self-dual (ASD) component, and the latter corresponds to BPS anti-M2 charge dissolved in the fluxes and a priori can be arbitrarily large. 

When the branes are smeared, this region, shown in grey in Figure \ref{fig:Setting}, is bounded in the infrared by a hypersurface at $\rho_1$, where the backreaction of the SD fluxes becomes dominant, and in the ultraviolet by a hypersurface at  $\rho_2$, where the anti-M2-dominated region is glued to the CGLP solution. 

To understand the origin of the hypersurface at $\rho_1$, we should remember that both in Polchinski-Strassler \cite{Polchinski:2000uf} and in mass-deformed M2 branes \cite{Bena:2000zb}, the naive pre-brane-polarization solution has an infrared singularity \cite{Girardello:1999bd,Pope:2003jp} which comes from the backreaction of the polarizing fluxes and which is excised by brane polarization. As we will argue in Appendix \ref{IRbackreaction}, when the harmonic function of the unpolarized branes goes like $Q/r^\Delta$, the backreaction of polarizing fields of strength $F$ modify it with a term of order $F^2 /r^{2 \Delta -2}$. This backreaction does not dominate the infrared for smeared anti-D3 branes in KS \cite{Bena:2012vz}, which have $\Delta=1$, and it clearly dominates the infrared of localized anti-M2 branes in CGLP, whose harmonic function will diverge as $1/r^6$. The story is more subtle for smeared anti-M2 branes, where $\Delta=2$ and therefore both the zeroth order warp factor and its $F^2$ correction have the same infrared growth. We have explicitly checked that, unlike for anti-D3 branes in KS, higher-order corrections in $F$ do give rise to more divergent terms. Nevertheless, in the brane-dominated region, these corrections to the warp factor are subleading compared to the fields of the anti-M2 brane.

In addition to the SD flux, the ASD flux of the supersymmetric anti-M2 CGLP solution can also cause infrared trouble. To see this, recall that the Stenzel BPS anti-M2 solution with ASD flux is very different from the BPS Stenzel M2 solution with SD flux constructed in \cite{Cvetic:2000db}, as the ASD flux either gives a singular infrared or a singular ultraviolet. Since we do not have the full solution it is hard to say precisely how much ASD flux we will have, and how strong it will be. However, we can estimate the strength of this flux and show that it does not affect the polarization potential.

To do this, we  first consider the $\widetilde{C}_1$ (infrared divergent) mode in Equation (\ref{anti-SDflux}). By analogy with (\ref{Maxwell-asympt}), the Maxwell charge in the deep IR (near the sources) and at infinity will be equal to:\footnote{As argued in \cite{Klebanov:2010qs}, the brane-flux annihilation of anti-M2 branes at the tip reduces the flux by two units and leaves behind $\left( \widetilde{M} - 1 - N_{\overline{\textrm{M2}}} \right)$ M2 brane sources. It is straightforward to see from (\ref{Maxwell-asympt}) and (\ref{Maxwell-asympt-anti}) that the asymptotic Maxwell charge remains the same during this process. There is a subtlety regarding the exact change in the flux in the brane/flux transition, but, as pointed out in the end of \cite{Klebanov:2010qs}, this concerns only $\mathcal{O}(1)$ quantities suppressed in the large $\widetilde M$ limit.} 
\begin{equation}
\label{Maxwell-asympt-anti}
Q^\textrm{Maxwell}_{\overline{\rm M2}} (0) = -N_{\overline{\textrm{M2}}}
\qquad \textrm{and} \qquad
Q^\textrm{Maxwell}_{\overline{\rm M2}} (\infty) = \dfrac{\widetilde{M}^2}{4}-N_{\overline{\textrm{M2}}} \, .
\end{equation}
We expect the full solution to differ in the UV from the CGLP one only by normalizable modes, which, as one can see from Equation (\ref{SDflux}) and the discussion that follows it, suggests that the flux functions, $f(\rho)$ and $h(\rho)$, will exponentially vanish in the UV. Comparing (\ref{Maxwell}) and  (\ref{Maxwell-asympt-anti}) we see that one can only reproduce the right charge in the UV if
$P \sim \widetilde{M}^2/4-N_{\overline{\textrm{M2}}}$. On the other hand, in order for the Maxwell charge near the anti-M2 branes to be $-N_{\overline{\textrm{M2}}}$, the flux functions in the IR must give an order $\widetilde{M}^2$ contribution to the Maxwell charge (\ref{Maxwell}):
\begin{equation}
\label{IR-flux-condition}
\dfrac{32 \pi^4}{\left( 2 \pi l_\textrm{P}\right)^6} \cdot h_\textrm{IR} (f_\textrm{IR} - 2 h_\textrm{IR}) \approx \dfrac{\widetilde{M}^2}{4} \, .
\end{equation}

To find the leading IR behavior of the flux functions one has to solve (\ref{sol-xif-xih-Expansion}) using the definitions of $\xi_f^-$ and $\xi_h^-$ in (\ref{xi-zfh}). The homogeneous part of the resulting solution is precisely the ASD flux of (\ref{anti-SDflux}), which dominates the small $\rho$ expansion of the fluxes. If one now substitutes $f(\rho) \approx 2 \left( - \widetilde{C}_1 + 4  \widetilde{C}_2 \right) + \ldots$ and $h(\rho) \approx \widetilde{C}_1 \rho^{-2}$ into the left hand side of (\ref{IR-flux-condition}) one finds that it can never be positive. This implies that a nonzero $\widetilde{C}_1$ gives a positive Maxwell charge near the anti-M2 branes, and indicates that $\widetilde{C}_1$ is very small. Thus, the IR-singular mode of the ASD flux does not affect the $\rho \gg \rho_1$ region of the solution.

The  $\widetilde{C}_2$ (UV divergent) mode of the ASD flux is a bit more tricky.  At the $\rho=\rho_2$ gluing hypersurface the ``incoming" CGLP solution has only SD fluxes, and one may hope that there will be no ASD flux on the anti-M2 dominated side and hence the effects of $\widetilde{C}_2$ on the polarization potential will be negligible. However, it may also happen that the nonlinearity of the equations of motion on the gluing surface will generate such a term and, while we cannot estimate its value in the absence of a fully-backreacted solution, it seems reasonable to assume that this flux will be of the same order as the incoming SD flux. When $\rho_2$ is large (which is the regime we have used in the previous section) the ASD mode proportional to $\widetilde{C}_2$ diverges exponentially and hence the only way to match the fluxes on the gluing hypersurface is if $\widetilde{C}_2$ is exponentially suppressed as $ M e^{-\rho_2}$. To summarize, both $\widetilde{C}_1$ and $\widetilde{C}_2$ are small and cannot impact our calculation in any way. This is precisely what we used in the last paragraph of Section~\ref{subsec:antiM2cglp}.

Another validity condition for our calculation is that the radius where the three terms in the polarization potential are detailed-balanced, $\rho_\star$, be smaller than $\rho_2$, such that polarization takes place inside the anti-M2 dominated region. An even more stringent condition, necessary if one is to be able to relate the transverse and the KP polarization potentials, is that $\rho_\star$ be smaller than the size of the large four-sphere, such that the polarization potential (\ref{Potential-final}) describes the physics in a region near the localized anti-branes where the un-warped geometry can be approximated by $\mathbb{R}^8$ and therefore the $AdS_4 \times S^7$ polarization analysis of \cite{Bena:2000zb} can be applied. 

It is easy to see that we can always make $\rho_\star$ small by considering the polarization potential of the anti-M2 branes into multiple M5. Indeed, as we discussed in section \ref{subsec:M2AdSpol}, the potential for all the anti-M2 branes to polarize into one M5 brane is given by simply replacing $n$ by the total number of anti-M2 branes, $N_{\overline{\textrm{M2}}}$, in equation (\ref{V-M5}). If one considers instead the polarization into $N_{\textrm M5}$ coincident M5 branes, each carrying $ N_{\overline{\textrm{M2}}} /N_{\textrm M5}$ units of anti-M2 charge, the full polarization potential is obtained by replacing $n$ with $ N_{\overline{\textrm{M2}}} /N_{\textrm M5}$ in equation (\ref{V-M5}), and multiplying the potential by an overall factor of $N_{\textrm M5}$. This will effectively lower $\rho_\star$ by a factor of $N_{\textrm M5}$, and will therefore always allow us to bring $\rho_\star$ within the desired range. Note that increasing the M5 charge of the polarization shell does not affect the $\rho^2$ term in the potential, and hence the conclusion that the anti-M2 branes are tachyonic is robust. 

The other important assumption we have made in obtaining the transverse-channel polarization potential of localized branes from that of smeared branes is that the $\rho^4$ and $\rho^2$ terms of that potential are independent of the position of the branes.  As we explained in Section \ref{subsec:M2AdSpol}, these terms can be related directly to non-normalizable modes in the ultraviolet of the brane-dominated region and, when one studies brane polarization in $AdS_4 \times S^7$  \cite{Bena:2000zb}, one fixes a-priori the values of these non-normalizable modes in terms of the mass-parameters of the dual theory. This ensures that the  $\rho^4$ and $\rho^2$ terms in the polarization potential are independent of the position of the branes.\footnote{The $\rho^6$ term comes  from the M5 branes of the polarizing shell wrapping a 3-sphere, and is independent by construction of the position of the anti-M2 branes.}

However, for anti-M2 branes in CGLP, the UV boundary conditions for the anti-M2 throat ``reside" at the $\rho=\rho_2$ hypersurface where this throat is glued to the asymptotically-CGLP solution of \cite{Cvetic:2000db}.  If the anti-M2 branes are localized on the $S^4$, this hypersurface will be deformed and will not be at constant $\rho$ any more. Hence, the boundary conditions for the closed and co-closed 4-form $Z^{-1} \left( \star_8 F_4 + F_4 \right)$ and for the $L=2$ modes that enter the $\rho^2$ terms will change, and therefore the polarization potential will be modified. 

To ensure that this effect is small we have to work in a region of parameters where $\rho_2$ is much larger than the distance over which the anti-M2 branes move, which is of order the size of the tip ($l_\epsilon = \epsilon^{3/4}$) and therefore the effect of moving the anti-M2 branes will be suppressed by a positive power of $l_\epsilon / \rho_2$. 

To see that one can always do this, one should first remember that our problem has only three free parameters:\footnote{For the CGLP solution with no sources ($N_{\textrm{M2},\overline{\textrm{M2}}}=0$) the $\epsilon$ parameter is not physical and can be gauged away. However, when branes are present, this parameter acquires a physical meaning, much like in Klebanov-Strassler \cite{oai:arXiv.org:hep-th/0511254}.} the CGLP magnetic flux, $\widetilde{M}$, the number of anti-M2 branes, $N_{\overline{\textrm{M2}}}$, and the size of the un-warped Stenzel tip, $l_\epsilon = \epsilon^{3/4}$. In the absence of supersymmetry-breaking, each of these parameters comes with its own scale: if one sets $\widetilde M$ to zero and considers (BPS) anti-M2 branes in a Stenzel space, the full solution will be warped $\mathbb{R}^{2,1}$ times Stenzel, with the warp factor given by the harmonic function sourced by the anti-M2 branes. This solution will be controlled by two scales: the Schwarzschild radius of the anti-M2 branes and $l_\epsilon$, the size of the Stenzel tip. Similarly, the BPS CGLP solution with BPS M2-branes is controlled by three parameters, $l_\epsilon$ and the ``Schwarzschild radii'' of the M2 branes and of the flux, which can be dialed at will.

\begin{figure}[t]
\centering
\includegraphics[scale=0.3]{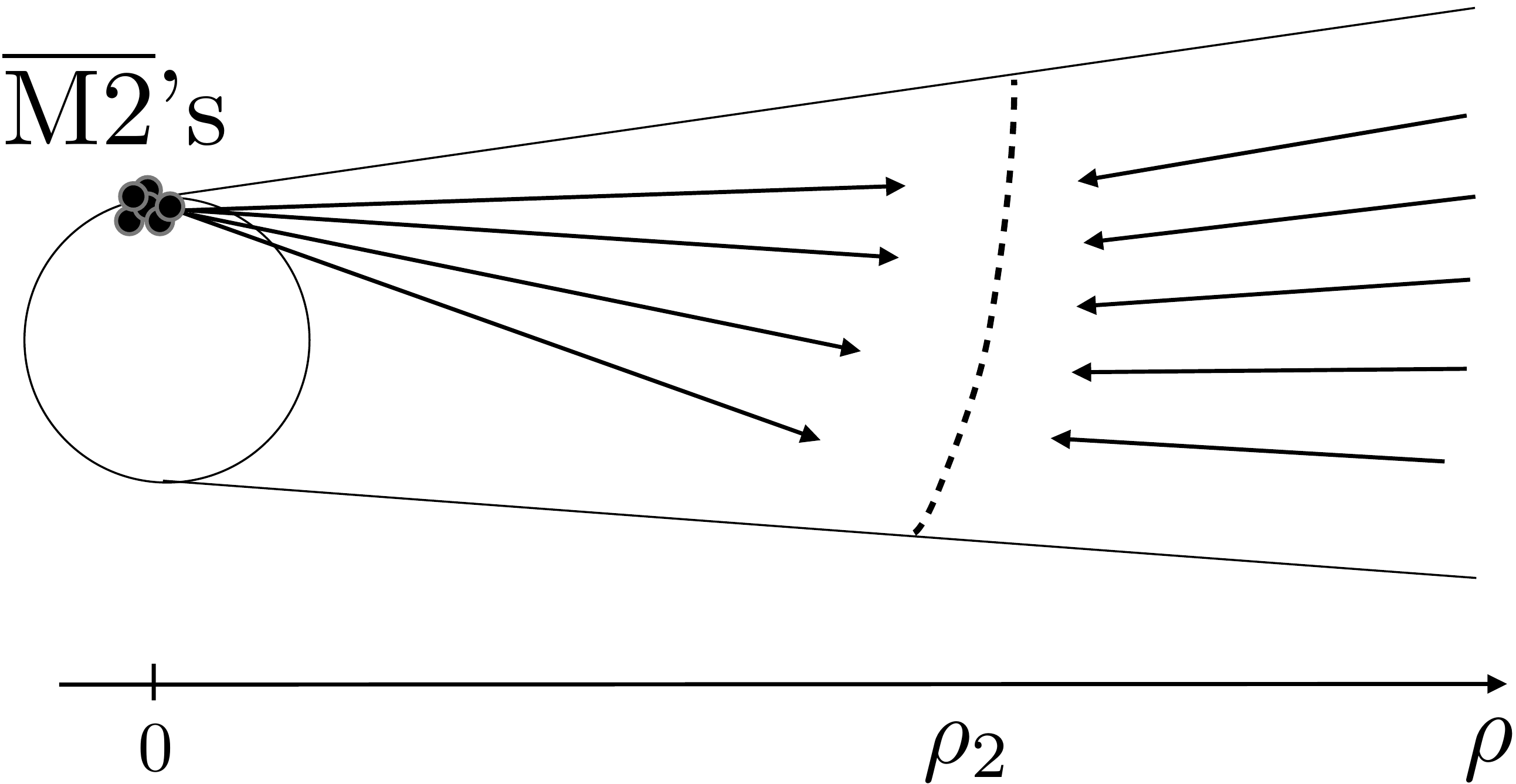}
\caption{For anti-M2 branes localized on the 4-sphere the gluing between the brane-dominated and the asymptotically-CGLP regions will not be at constant $\rho_2$. However, we can always push this hypersurface away from the tip (arrows pointing right) by increasing $N_{\overline{\textrm{M2}}}$, and push it towards the tip (arrows pointing left) by increasing $M$. If the size of the tip is much smaller than $\rho_2$, then localizing the branes will not affect this surface.}
\label{fig:r2}
\end{figure}

Even if we do not have the exact fully-back-reacted anti-M2 solution, it is clear that the position of the hypersurface where the anti-M2 region is glued to CGLP, $\rho_2$, is determined by a balance between the anti-M2 branes and the CGLP flux, $M$, and that increasing the number of anti-M2 branes pushes this surfaces to larger values of $\rho_2$. This can be done while still keeping 
$N_{\overline{\textrm{M2}}} \ll M^2$ such that the charge at infinity remains positive. The situation is shown on Figure \ref{fig:r2}. On the other hand, the size of the tip, $l_{\epsilon}$, will not enter in this balance, and therefore can be set to be much smaller than $\rho_2$. This ensures that the physics at the gluing surface is not affected by moving the anti-M2 branes at the tip, and hence that the smeared and localized polarization potentials in the transverse channel are the same. 

It is important to note that one can go to the regime of parameters where $\rho_2 > l_\epsilon$ without affecting the other assumptions we made about the polarization radius $\rho_\star$ and the IR cut-off $\rho_1$.

\section{Conclusions} 
\label{sec:concl}

We have studied the dynamics of anti-M2 branes placed at the bottom of a supersymmetric background with M2 brane charge dissolved in flux (the CGLP solution of~\cite{Cvetic:2000db}) taking into account the full backreaction of the anti-branes on the ambient geometry. We proved that the anti-M2 solution has a singularity in the energy density of the four-form flux, confirming the linearised analysis of~\cite{Bena:2010gs}. We then looked for a resolution of such singularity by brane polarization, as suggested by the probe picture of Klebanov and Pufu~\cite{Klebanov:2010qs}, in which the anti-M2 branes expand into one or more M5 branes wrapping an $S^3$ inside the $S^4$ in the infrared.  Since our starting point was a solution for anti-M2 branes smeared on the $S^4$, we could not compute directly the polarization potential for the Klebanov-Pufu channel, so we first computed the potential for polarization into M5 branes wrapping the shrinking $S^3$ of the CGLP geometry, at a finite distance from the tip. We found that the potential has no minimum away from the tip, signaling that the breaking of supersymmetry alters qualitatively the phase structure of the supersymmetric M2-M5 polarization~\cite{Bena:2000zb}. This happens because of the contribution of $L=2$ modes, corresponding to traceless boson bilinears in the world-volume of the anti-M2's, which break supersymmetry. 

We then argued that, at least in some region of the parameter space, the polarization potential for smeared anti-M2 branes is not sensitive to the position of the sources, and is thus the same as the potential for localized anti-M2 branes expanding into M5 branes on the shrinking $S^3$. This in turn allowed us to extract the $L=2$ mode that enters into the $r^2$ term of the polarization potential of localized anti-M2 branes and thus to explicitly compute the potential for polarization of localized back-reacted anti-M2 branes in the Klebanov-Pufu channel. To our great surprise, we found that this potential has an $\theta^2$ term that is never positive. Since this term is the same as the force acting on a mobile anti-M2 brane placed in the background sourced by the other anti-M2 branes, this implies that anti-M2 branes at the bottom of the CGLP background repel each other, and hence that the world-volume theory of CGLP anti-M2 branes has a tachyon. This in turn would imply that the putative $AdS_4 \times S^7$ throat sourced by anti-M2 branes localized at the bottom of CGLP background would be unstable to fragmentation. 

\begin{figure}[t]
\centering
\includegraphics[scale=0.50]{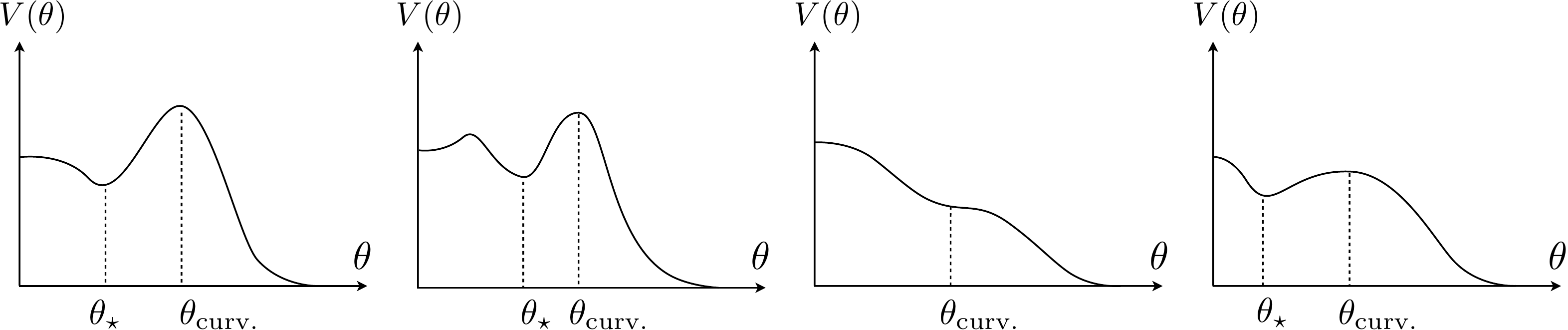}
\caption{The naive polarization potential one derives ignoring the anti-M2 backreaction \cite{Klebanov:2010qs} (\emph{first} graph), the potential one obtains by (incorrectly) assuming that backreaction will give rise to an attractive force between the anti-branes (\emph{second} graph) and the two possible corrected potentials obtained by including the anti-M2 tachyon. If $\theta_\star$ is larger than $\theta_\textrm{curv}$, than the tachyonic $-\theta^2$ mode can wipe out the local minimum (\emph{third} graph). We can reduce $\theta_\star$ by considering polarization into multiple M5 branes, guaranteeing this way a metastable minimum at the $\theta=\theta_\star$ scale (\emph{fourth} graph). As we explain in the text, the tachyonic mode can greatly change the physics of this metastable vacuum, by opening up new instability directions.}
\label{fig:kpv3}
\end{figure}

It is very important to understand the end-point of this tachyonic instability. As we show in Figure \ref{fig:kpv3}, the brane-flux annihilation potential comes with two scales: $\theta_\star$, where the three terms in the North-Pole potential (\ref{locpotential}) are detailed balanced, and $\theta_{\rm curv}$, where the curvature of the sphere begins to have an important effect and pulls the M5 branes over the equator triggering this way brane-flux annihilation. Clearly when $\theta_\star$ is larger than $\theta_{\rm curv}$ there is no metastable minimum, and the branes undergo immediate brane-flux annihilation. This happens for example in \cite{Klebanov:2010qs} when the number of anti-M2 branes polarizing into a shell with M5 dipole charge one is larger than $5.4\%$ of $M$.

Our analysis tells us that anti-M2 branes repel at distances smaller than the size of the four-sphere, and hence that the brane-flux annihilation potential differs from the one calculated using the probe approximation \cite{Klebanov:2010qs} by a negative contribution. Indeed, if one ignores backreaction, the North Pole expansion of the probe potential  begins with a negative term of order $\theta^4$ \cite{Klebanov:2010qs}, while we find that this expansion should begin instead with a negative term of order $\theta^2$. Still, since we only trust our calculations at the scale $\theta_\star$, we cannot say what is the functional form of this negative contribution at the scale $\theta_{\rm curv}$, and hence we cannot determine conclusively whether the metastable vacuum with all anti-M2 branes polarized into one M5 brane gets destabilized or not.

Nonetheless, M2 branes can also polarize into multiple M5 branes and, as we explained in Section \ref{subsec:M2AdSpol}, this reduces $\theta_\star$ by the square root of the number of M5 branes. Hence, by increasing the M5 dipole charge of the polarization shell we can arrange for $\theta_\star$ to become much smaller than $\theta_{\rm curv}$. Since at this scale the polarization potential has three terms and since the $\theta^2$ term is negative and the $\theta^6$ term is positive, this guarantees that this potential will have a metastable minimum. Hence, the tachyon we find, while potentially-dangerous for the metastable vacua with small M5 dipole charge, will not wipe out the metastable vacua that have a large M5 dipole charge. 

However, the presence of the tachyon does not bode well for the stability of such vacua. Indeed, the tachyon of the M2 branes translates into a tachyon for the two-form field on the M5 worldvolume, which 
indicates that the whole configuration is unstable. Furthermore, if one calculates the potential for one anti-M2 brane to shoot out of the polarization shell, the near-shell solution is dominated by the anti-M2 branes, and therefore this potential will be repulsive. 

The metastable shell with a large M5 dipole charge can also decay by peeling out M5 shells that brane-flux annihilate. To see how this can happen, consider the fluctuation where one of the M5 branes gets more anti-M2 charge than its friends. Since shells with larger anti-M2 charges have larger equilibrium radii (\ref{rhostar}), this overcharged M5 brane will be driven to a larger value of $\theta$. 
Furthermore, as the anti-M2 branes in its world-volume repel the anti-M2 branes in the remaining M5 branes, the force driving it to larger values of $\theta$ increases, and this M5 brane will therefore have the tendency to slide over the equator triggering brane-flux annihilation\footnote{Note that this would not happen if antibranes attracted; as argued in \cite{DeWolfe:2004qx} this attraction would push multiple nearby shells to merge and form a single shell.}.

Thus, our analysis indicates that there exist metastable back-reacted anti-M2 shells with a large M5 dipole charge. Nevertheless, the presence of an anti-M2 tachyon makes the physics of these vacua very different than the one expected from the probe approximation. These shells can decay by shooting out anti-M2 branes, by rapid brane-flux annihilation caused  by the peeling out of charged M5 shells, as well as by the fragmentation of the $AdS_4 \times S^7$ throat sourced by the anti-M2 branes. 

Perhaps the most important question our calculation raises is whether the tachyon we find is just an accidental feature of anti-M2 branes in CGLP, or is a more generic characteristic of all anti-branes in backgrounds of opposite charge. The fact that anti-brane singularities do not appear to be cloakable by regular event horizons~\cite{Bena:2012ek,Bena:2013hr,Buchel:2013dla} points towards the latter option. 

It would be very exciting if one could extend our calculation and establish whether anti-D3 branes in Klebanov-Strassler~\cite{Kachru:2002gs} are also tachyonic. Their back-reacted polarization potential in the transverse channel was calculated in~\cite{Bena:2012vz}, both for smeared and for localized branes, and it was found that this potential has no metastable minimum. However, extracting the back-reacted KPV polarization potential from the transverse one is not as straightforward as for anti-M2 branes, essentially because, as we explained in Section \ref{subsec:reviewPS}, the Polchinski-Strassler polarization potential can have {\em two} supersymmetry-breaking terms compatible with the symmetry of the problem: an $L=2$ mode and a gaugino mass. Since the second term is absent for anti-M2 branes, knowing the terms of the M2 transverse polarization potential had enough information to allow us to calculate the other polarization potential. However, to do this for anti-D3 branes in KS one needs first to disentangle the two supersymmetry-breaking contributions, which is more subtle. We plan to report on this in upcoming work.

Notice that the anti-brane repulsion we found seems to contradict the result of  \cite{DeWolfe:2004qx}, where it was argued that branes in backgrounds of opposite charge should attract (see also \cite{Blaback:2012nf}). The intuition behind this claim was that anti-branes should create a cloud of flux of opposite charge around them, which in turn screens their negative charge making the electromagnetic repulsion weaker than the gravitational attraction. Our explicit calculations fail to see such a screening effect, but rather show that the cloud sourced by the anti-branes would have more charge than mass, and hence repel other anti-branes. Of course, since our calculations are valid in the regime when the inter-brane separation is smaller than the size of the cloud sourced by the anti-branes, they test some very non-linear dynamics, which the arguments of \cite{DeWolfe:2004qx,Blaback:2012nf}, valid in the regime where the inter-brane distance is larger than the size of cloud, do not take into account. 

It would be very interesting to try to reproduce our tachyonic instability via a more direct first-principles calculation, and to see whether it is also exists in different regimes of parameters than the one we consider here. If this tachyonic mode is generic, it would point towards a very basic feature of the interaction between branes and fluxes of opposite charge.

\acknowledgments{We would like to thank Matteo Bertolini and Thomas Van Riet for useful discussions. I.B. and S.K. are supported in part by the ANR grant 08-JCJC-0001-0 and by the ERC Starting Grant 240210 -- String-QCD-BH. I.B. is also supported by a grant from the Foundational Questions Institute (FQXi) Fund, a donor advised fund of the Silicon Valley Community Foundation on the basis of proposal FQXi-RFP3-1321 to the Foundational Questions Institute. This grant was administered by Theiss Research. The work of M.G. and S.K. is supported in part by the ERC Starting Grant 259133 -- ObservableString. The work of S.M. is supported by the ERC Advanced Grant 32004 -- Strings and Gravity.}

\appendix

\section{Explicit form of the \texorpdfstring{$\xi_a$}{xi} equations}
\label{sec:xi's}

Here we summarize the expressions for $\xi_a$'s that follow from their definition in~\eqref{FirstOrderSUSY} and the explicit form of the superpotential and the kinetic term in~\eqref{superpotential} and~\eqref{G-metric} respectively.

For the metric functions $\alpha$, $\beta$ and $\gamma$ we have:
\begin{eqnarray}
\xi_\alpha^+ = \xi_\alpha^- &=& 3 e^{2 \alpha + 4 \beta} - 3 \left( 2 {\alpha^\prime} + 3 {\beta^\prime} + {\gamma^\prime} \right)  e^{3 ( \alpha + \beta )} + 3 e^{2 ( \alpha + \beta + \gamma )} + 6 e^{4 \alpha + 2 \beta}  \, ,
\nonumber
\\
\xi_\beta^+ = \xi_\beta^- &=& 6 e^{2 \alpha + 4 \beta} - 3 \left( 3 {\alpha^\prime} + 2 {\beta^\prime} + {\gamma^\prime} \right)  e^{3 ( \alpha + \beta )} + 3 e^{2 ( \alpha + \beta + \gamma )} + 3 e^{4 \alpha + 2 \beta} \, ,
\nonumber
\\
\xi_\gamma^+ = \xi_\gamma^- &=& - 3 \left( {\alpha^\prime} + {\beta^\prime} \right)  e^{3 ( \alpha + \beta )} + 3 e^{2 ( \alpha + \beta + \gamma )}  \, ,\label{xi-abg}
\end{eqnarray}
while the dual modes for the warp function and the fluxes are given by:
\begin{eqnarray}
\label{xi-zfh}
\xi_z^+ &=& \frac{9}{2} e^{3(\alpha + \beta)} \left( z^\prime - \frac{1}{3} e^{3 z} K^\prime \right) \, ,
\nonumber
\\
\xi_f^+ &=&\frac{1}{2} e^{3(\alpha - \beta - z)} \left( {f^\prime} + 6 e^{3(\beta - \alpha)} h \right) \, ,
\nonumber
\\
\xi_h^+ &=& 6 e^{- \alpha + \beta - 3 z} \left( {h^\prime} + \frac{1}{2} e^{\alpha - \beta} (f -4 h) \right) \, ,
\nonumber
\\
\xi_z^- &=& \frac{9}{2} e^{3(\alpha + \beta)} \left( z^\prime + \frac{1}{3} e^{3 z} K^\prime \right) \, ,
\\
\xi_f^- &=& \frac{1}{2} e^{3(\alpha - \beta - z)} \left( {f^\prime} - 6 e^{3(\beta - \alpha)} h \right) \, ,
\nonumber
\\
\xi_h^- &=& 6 e^{- \alpha + \beta - 3 z} \left( {h^\prime} - \frac{1}{2} e^{\alpha - \beta} (f -4 h) \right) \, .
\nonumber
\end{eqnarray}

\section{Explicit form of the \texorpdfstring{$\xi_a^\prime$}{xi-prime} equations}
\label{sec:xi's-dot}

The first-order equations of motion for the modes $\xi_a^\pm$ follow directly from (\ref{SecondOrder}), from the explicit form of the metric $G_{ab}$, and from the superpotentials in (\ref{G-metric}) and (\ref{superpotential}) respectively. To find the polarization potential we only needed the equations for the flux modes and for the warp function (\ref{SecondOrder}). Here we give for completeness the remaining equations. Together with (\ref{SecondOrder}) and the definition of the $\xi_a$ modes (\ref{xi-abg}), (\ref{xi-zfh}), they form a system of twelve first-order differential equations which is equivalent to the equations of motion derived from the one dimensional action defined by (\ref{G-metric}) and (\ref{superpotential}). The remaining three ${\xi_a^-}^\prime$ equations are:
\begin{eqnarray}
\label{xi-abc-dot} 
{\xi_\alpha^-}^\prime + {\xi_\beta^-}^\prime  &=& \frac{1}{2} e^{-3 \left(\alpha+\beta \right)} \left( {\xi_\alpha^-}^2 + {\xi_\beta^-}^2 + 5 {\xi_\gamma^-}^2 + \frac{4}{3} {\xi_z^-}^2 \right) - e^{ -\alpha-\beta+2 \gamma} \left( \xi_\alpha^- + \xi_\beta^- + 5 \xi_\gamma^- \right) 
\nonumber \\
 && \qquad 
 + \frac{2}{3} e^{-3 \left(\alpha+\beta \right)} \xi_z^- \left( \xi_z^- - 3 e^{ 3 \left( \alpha+\beta+z \right) } K^\prime \right) \, ,
\nonumber \\ 
{\xi_\alpha^-}^\prime - {\xi_\beta^-}^\prime  &=& \left(2 e^{\alpha-\beta} - e^{-\alpha+\beta} \right) \xi_\alpha^- + \left(e^{\alpha-\beta} - 2 e^{-\alpha+\beta} \right) \xi_\beta^- + 36 h e^{3 \left(-\alpha+\beta \right)} \xi_f^-
\nonumber \\
 && \qquad 
  - (f-4h) e^{\alpha-\beta } \xi_h^- + \left( 6 e^{3 \left( - \alpha + \beta + z \right)} {\xi_f^-}^2 - \frac{1}{6} e^{ \alpha - \beta + 3 z } {\xi_h^-}^2\right)  \, ,\nonumber
\\
{\xi_\gamma^-}^\prime  &=&  e^{- \alpha - \beta + 2 \gamma} \left( \xi_\alpha^- + \xi_\beta^- + 5 \xi_\gamma^-  \right) \, .
\end{eqnarray} 
The six equations for ${\xi_a^+}^\prime$ are:
\begin{eqnarray}
\label{xi-abg-dot-PLUS}
{\xi_\alpha^+}^\prime + {\xi_\beta^+}^\prime  &=& \frac{1}{2} e^{-3 \left(\alpha+\beta \right)} \left( {\xi_\alpha^+}^2 + {\xi_\beta^+}^2 + 5 {\xi_\gamma^+}^2 + \frac{4}{3} {\xi_z^+}^2 \right) - e^{ -\alpha-\beta+2 \gamma} \left( \xi_\alpha^+ + \xi_\beta^+ + 5 \xi_\gamma^+ \right) 
\nonumber \\
 && \qquad 
 + \frac{2}{3} e^{-3 \left(\alpha+\beta \right)} \xi_z^+ \left( \xi_z^+ + 3 e^{3 \left( \alpha+\beta+z \right)} K^\prime \right) \, ,
\nonumber \\ 
{\xi_\alpha^+}^\prime - {\xi_\beta^+}^\prime  &=& \left(2 e^{\alpha-\beta} - e^{-\alpha+\beta} \right) \xi_\alpha^+ + \left(e^{\alpha-\beta} - 2 e^{-\alpha+\beta} \right) \xi_\beta^+ - 36 h e^{3 \left(-\alpha+\beta \right)} \xi_f^+
\nonumber \\
 && \qquad 
  + (f-4h) e^{\alpha-\beta } \xi_h^+ + \left( 6 e^{3 \left( - \alpha + \beta + z \right)} {\xi_f^+}^2 - \frac{1}{6} e^{ \alpha - \beta + 3 z } {\xi_h^+}^2\right)  \, ,\nonumber
\\
{\xi_\gamma^+}^\prime  &=&  e^{- \alpha - \beta + 2 \gamma} \left( \xi_\alpha^+ + \xi_\beta^+ + 5 \xi_\gamma^+  \right) \, 
\end{eqnarray}
and
\begin{eqnarray}
\label{xi-fhz-dot-PLUS}
{\xi^+_f}^\prime &=& - 2 e^{-3 (\alpha+\beta+z)} h \, \xi^+_z + \frac{1}{2} e^{\alpha-\beta}  \xi_h^+ \, ,
\nonumber \\
{\xi^+_h}^\prime &=& - 2 e^{-3 (\alpha+\beta+z)} (f-4h) \, \xi^+_z + 6 e^{3(\beta-\alpha)}  \xi_f^+ - 2 e^{\alpha-\beta}  \xi_h^+ \, ,
\\
{\xi^+_z}^\prime &=&  K^\prime e^{3 z} \xi^+_z - \frac{e^{3 z}}{4} \left( 12 e^{3 (\beta-\alpha)} {\xi_f^+}^2  +  e^{\alpha-\beta} {\xi_h^+}^2 \right)  
 \nonumber \, .
\end{eqnarray}

\section{The infrared backreaction of the polarizing fields}
\label{IRbackreaction}

In this appendix we give a simple and intuitive argument that allows us to find the infrared divergence caused by the backreaction of the forms that trigger brane polarization on the metric warp factor. We also show why the far infrared region is so different than in the smeared anti-D3 setup, giving a brane-dominated region that does not extend all the way to $\rho=0$.

As we discussed in great detail above, one has to  exclude the $\rho \ll \rho_1$ part of geometry from the brane dominated region, since for small $\rho$ the flux becomes singular. For anti-D3's smeared over the blown-up 3-sphere of the warped conifold, we also have to allow flux singularity in the IR, since otherwise the flux remains IASD all the way to the UV \cite{Bena:2012bk}. The situation there, nevertheless, differs from the setup discussed in this paper. First, the IASD flux on the conifold is the same as the ISD one up to a sign of the $B$-field (see the end of Section~\ref{subsubsec:ASD}), and so it can be regular both in the UV and in the IR of the anti-D3 throat. Second, the GPPZ-like singularity of the ISD flux in the throat is not strong enough to distort the leading order behavior of the warp function. To be more precise, it produces only a $\rho^{-1}$ correction to the warp function, which is subleading to the un-perturbed $\rho^{-2}$ solution. 

Let us provide a simple intuitive argument for that statement. The quadratic term in the polarization potential is given by the force felt by a probe brane in the perturbed throat geometry. In our setup it is given by $e^{-3 z} - K$, see (\ref{Potential-not-final}). For the KS setup, $e^{3 z}$ is replaced by the D3 warp function $Z$, and $K$ becomes the 5-form flux.  At the zeroth order in the flux expansion $Z^{-1}_0 - K_0$ vanishes. For the sake of generality, let us set $Z^{-1}_0, K_0 \sim \rho^{\Delta}$ for small $\rho$. We will also assume for simplicity that  for the (second order) perturbed solution $\delta Z \sim \mathcal{M}_\textrm{SD}^2 \rho^{-\Delta+\delta}$ and $\delta K \sim \delta Z$.

Plugging this into $Z^{-1} - K$ and expanding to the $\mathcal{M}_\textrm{SD}^2$ order, we see that the first non-zero contribution is of order $\rho^{\Delta+\delta}$. Therefore, to obtain an $\rho^2$ term in the potential, we need:
\begin{equation}
\label{delta}
\delta = 2- \Delta  \, .
\end{equation}
For anti-D3's smeared over the 3-cycle of the deformed conifold $\Delta=1$, and so $\delta=1$. We see that the perturbed warp function is small compared to zeroth order one. It means that the flux singularity in this case is not sufficiently strong to modify the $\rho^{-2}$ behavior of the warp function near the source. We learn that the flux perturbation never dominates in the deep IR.

For our anti-M2 configuration $\Delta=2$ implying $\delta=0$. Thus the perturbation now has exactly the same near source behavior as the unperturbed warp function, and this is the reason why we have to exclude the $\rho \ll \rho_1$ region, where $\mathcal{M}_\textrm{SD}$ is not sufficiently small to trust the expansion.

Let us also mention that the simple formula (\ref{delta}) also reproduces correctly the result for localized D3 branes. In this case $\Delta = 4$ and $\delta = -2$. Thus, contrary to the smeared case, for localized D3 branes the flux perturbation completely destroys the brane throat in the IR, explaining the naked singularity of the GPPZ flow~\cite{Girardello:1999bd}. The solution  has been explicitly computed to the second order in the flux perturbation by Freedman and Minahan in~\cite{Freedman:2000xb}. Their result is\footnote{Here the radial coordinate is denoted by $r$, see footnote \ref{r-rho}.} a $M^2 r^{-6}$ contribution to the warp factor, in nice agreement with our expectations.

\bibliographystyle{utphys}
\bibliography{Final-M2}

\end{document}